\documentclass[usenatbib]{mn2e}
\newcommand{\bib}{\bibitem[\protect\citeauthoryear}
\usepackage{epsf}
\usepackage{times}
\begin{document}
\title[Backflow around radio jets]{Relativistic jet models for two  
  low-luminosity radio galaxies: evidence for backflow?}
\author[R.A. Laing \& A.H. Bridle]
   {R.A. Laing \thanks{E-mail: rlaing@eso.org}$^{1}$, A.H. Bridle $^2$\\
    $^1$ European Southern Observatory, Karl-Schwarzschild-Stra\ss e 2, D-85748 
    Garching-bei-M\"unchen, Germany \\ 
    $^2$ National Radio Astronomy Observatory, Edgemont Road, Charlottesville,
    VA 22903-2475, U.S.A. \\
}

\date{Received }
\maketitle

\begin{abstract}
We show that asymmetries in total intensity and linear polarization between
the radio jets and counter-jets in two lobed Fanaroff-Riley Class I (FR\,I)
radio galaxies, B2\,0206+35 (UGC\,1651) and B2\,0755+37 (NGC\,2484), can
be accounted for if these jets are intrinsically symmetrical, with decelerating relativistic
outflows surrounded by mildly relativistic backflows.  Our
interpretation is motivated by sensitive, well-resolved Very Large Array imaging
which shows that both jets in both sources have a two-component
structure transverse to their axes. Close to the jet axis, a centrally-darkened counter-jet
lies opposite a centrally-brightened jet, 
but both are surrounded by broader collimated emission that is brighter
on the counter-jet side. We have adapted our previous models
of FR\,I jets as relativistic outflows to include an added component
of symmetric backflow. We find that the observed radio emission, after
subtracting contributions from the extended lobes, is well described by models in
which decelerating outflows with parameters similar to those derived for jets 
in plumed FR\,I sources are surrounded by backflows containing predominantly
toroidal magnetic fields. These return to within a few kpc of the galaxies with
velocities $\approx 0.25c$ and radiate with a synchrotron spectral index $\alpha \approx 0.55$.
We discuss whether such backflow is to be expected in lobed FR\,I sources and suggest
ways in which our hypothesis can be tested by
further observations.
\end{abstract}


\section{Introduction}
\label{Introduction}

Relativistic jet outflows from radio galaxies are a primary mechanism for
energy extraction from supermassive black holes in active galactic nuclei
(AGN) and an important source of
energy input to the intergalactic medium (IGM) in groups and clusters
(e.g. \citealt[and references therein]{McN07}).  
We are studying relativistic jet kinematics and dynamics in nearby
low-luminosity radio galaxies with Fanaroff-Riley Class I (FR\,I -
\citealt{FR74}) morphology for which we have obtained radio imaging and
polarimetry at high angular resolution transverse to the jets as well as along
their lengths.  
We have developed procedures
 for deriving three-dimensional variations of
intrinsic jet parameters -- velocity field, emissivity  and
magnetic-field ordering -- from an analysis of {\sl systematic} asymmetries
between the jets and  counter-jets \citep{LB02a,CL,CLBC,LCBH06}.  We compare the observed asymmetries in
images of total intensity, degree of linear
polarization and apparent magnetic field direction with the predicted effects of
relativistic aberration on synchrotron emission from particles in
partially-ordered magnetic fields in model outflows and deduce the
distributions of intrinsic properties within the jets.  We have found that a
generic property of the jet outflows in FR\,I radio galaxies is that they
decelerate from relativistic speeds ($\beta = v/c \approx 0.8$ -- 0.9) near the AGN to
subrelativistic speeds a few kiloparsecs away, and that the outflows are
systematically faster on-axis than at their edges.

It is critical for such an analysis to distinguish patterns of asymmetry in the jets
produced by relativistic aberration from any that are intrinsic to the
outflows or
which result from interactions between the outflows and anisotropic
environments, e.g.\ from pressure gradients or winds in the IGM.  One asymmetry
in FR\,I radio jets that has proven instructive in some sources and
problematic in others is the {\sl systematic difference between transverse
intensity profiles in the brighter jets and weaker counter-jets} when
observed at high sensitivity and angular resolution.

This difference correlates with indicators of the
orientation of the jets to the line of sight.
A statistical study of FR\,I jets in the B2 sample by
\cite{LPdRF} found
that the ratio of jet to counter-jet FWHM measured by Gaussian fitting at the
same distance from the nucleus on both sides is strongly anticorrelated with the average
jet/counter-jet brightness ratio
and with the ratio of core\footnote{The 'core' is defined as an unresolved
  component coincident with the AGN. The core/extended flux-density ratio is
  a statistical indicator of orientation.} to extended flux density.

This anticorrelation is qualitatively as expected for intrinsically symmetrical
relativistic outflows which are faster on-axis than at their edges.  In
this case, relativistic aberration makes the transverse brightness profiles of the
approaching, hence apparently brighter, jet more centrally peaked than those
of the receding counter-jet.  Gaussian fitting to the jet and counter-jet FWHM
then yields smaller values of the width for the apparently brighter jets even if
the (slower moving) outer boundaries of the jets appear identical on both sides
of the AGN.

The amplitude of the effect found in the B2 source sample by
\cite{LPdRF} is, however, surprisingly large.  Modelling of the
anticorrelation requires that the velocity $\beta_{\rm on-axis}
\approx 0.7$ and $\beta_{\rm edge} \approx 0.1$ \citep{LPdRF} in order
to reproduce the spread of width ratios. Two lines of argument suggest
that such large velocity ratios are not typical of the FR\,I
population. Firstly, the ratio $\beta_{\rm edge} / \beta_{\rm
  on-axis}$ required to explain the effect is quantitatively
inconsistent with the brightness and polarization distributions in
four of the five individual FR\,I sources we have modelled
(\citealt{LB02a,CL,CLBC} -- the exception is 3C\,296;
\citealt{LCBH06}). Secondly, the smallest values of $\beta_{\rm edge}
/ \beta_{\rm on-axis}$ are required only to generate the unusually
small values of jet/counter-jet width ratio $\approx 0.6$ in a few
members of the B2 source sample with particularly high jet/counter-jet
brightness ratios, whose jets are thought to be highly inclined to the
plane of the sky \citep{LPdRF}.

Thus far, our results would be consistent with the idea that all FR\,I jets are
symmetrical outflows, but that only a few have very large transverse velocity
gradients. Even this hypothesis fails for two of the B2 sample members, B2\,0206+35
and B2\,0755+37 \citep{LGBPB}\footnote{From now on we drop the B2.}. These
sources are unusual in that the {\sl lower} isophotes of their brighter jets
{\sl also} appear narrower than those of the counter-jets at the same distance
from the AGN in images of moderate resolution and sensitivity
(e.g.\ \citealt{Bondi00}) - even though the jets clearly exhibit the basal
asymmetries associated with symmetrical decelerating relativistic outflows.
Apparent width asymmetry in the fainter jet emission cannot generally be
explained by relativistic effects alone if the jets are both {\em symmetrical}
and {\em purely outflowing}\footnote{We discuss a special magnetic-field
  configuration for which this is not the case in Appendix~\ref{toroidal}.}.  On
the other hand, if the asymmetry is attributed to intrinsic or environmental
differences on the two sides of the AGN (e.g.\ \citealt{Bondi00}) there should
be no systematic trend for the wider jet to be on the receding side as it is in
the (albeit small) sample of \citet{LPdRF}.

In this paper, we explore an alternative explanation for the transverse
brightness profile asymmetries of the jets and counter-jets in 0206+35 and
0755+37.  This work was motivated by new deep imaging of these sources showing: 
(a) that their counter-jets have minima in their emission profiles with the same
widths as the main jets at similar distances from the nucleus and (b) that the
main jets are surrounded by faint emission resembling the broader outer
emission in  the counter-jets \citep[and Section~\ref{images}, below]{LGBPB}.
The new imaging data lead us to model the jets in these sources as intrinsically
symmetrical outflows near the jet axis surrounded by broader features from {\sl
  backflowing} material.  If backflow in the broader features can be
approximately symmetrical and mildly relativistic, then aberration
can make its emission appear slightly brighter on the {\sl counter-jet}
side, producing differences in isophotal width between the jets similar to those
observed.

Backflow is a reasonable hypothesis a priori for FR\,I sources like
0206+35 and 0755+37 {\sl whose jets appear to propagate within well-defined
lobes}.
It has been an acknowledged ingredient of models of lobed FR\,II
sources since the first attempts to simulate their hydrodynamics
\citep{Norman82}.
FR\,I sources cannot form lobes without similar deflection of jet material and
\cite{LGBPB} showed that FR\,I lobes resemble those of FR\,II sources in many
respects.  If FR\,I jets are much lighter than their surroundings and
initially fast 
(e.g.\ \citealt{LB02b}), we should not be surprised if some large-scale
post-jet flow in 
FR\,I lobes is marginally relativistic.  We also note that mildly relativistic
backflow
extends almost all the way back
to the centre of the host galaxy in simulations of relativistic FR\,I jets
with initial dynamical
flow parameters matching those deduced from our observations of 3C\,31 and
realistic pressure and density profiles for the surrounding IGM 
\citep{LB02b,PM07}.

In this paper, we show that a fully symmetrical model in which a decelerating
axisymmetric outflow is surrounded by a slower (but still slightly
relativistic) backflow
is {\sl quantitatively} consistent with the detailed brightness and
polarization distributions
of the jets and counter-jets in 0206+35 and 0755+37.  It is not obvious a
priori 
that conditions needed to produce {\sl symmetrical} backflow are likely to be
realised in lobed FR\,I radio galaxies. Nevertheless, our results suggest that
mildly relativistic backflow contributes significantly to the observed jet vs
counter-jet width relationships and we suggest ways in which this (perhaps
unexpected) ingredient
of FR\,I source structure could be investigated further.

In Section~\ref{obs}, we summarize the optical and large-scale radio properties
of the sources and discuss the additional image processing required to separate
jet and lobe emission.  Section~\ref{fits} describes our modelling procedure and
Section~\ref{compare} gives a comparison between models and data. The model
parameters are presented in Section~\ref{intrinsic}. A brief discussion is given
in Section~\ref{discuss}.  Section~\ref{summary-further} summarizes our
conclusions and suggests further work. Finally, Appendix~\ref{toroidal} demonstrates 
that a toroidally-magnetized outflow can, in special circumstances, produce jet/counter-jet 
sidedness ratios significantly less than unity.
 
We adopt a concordance cosmology with Hubble constant, $H_0$ =
70\,$\rm{km\,s^{-1}\,Mpc^{-1}}$, $\Omega_\Lambda = 0.7$ and $\Omega_M =
0.3$.

\section{Observations and images}
\label{obs}

\subsection{The sources: optical data and large-scale radio structures}
\label{sources}

\begin{figure}
\begin{center}
\epsfxsize=8.5cm
\epsffile{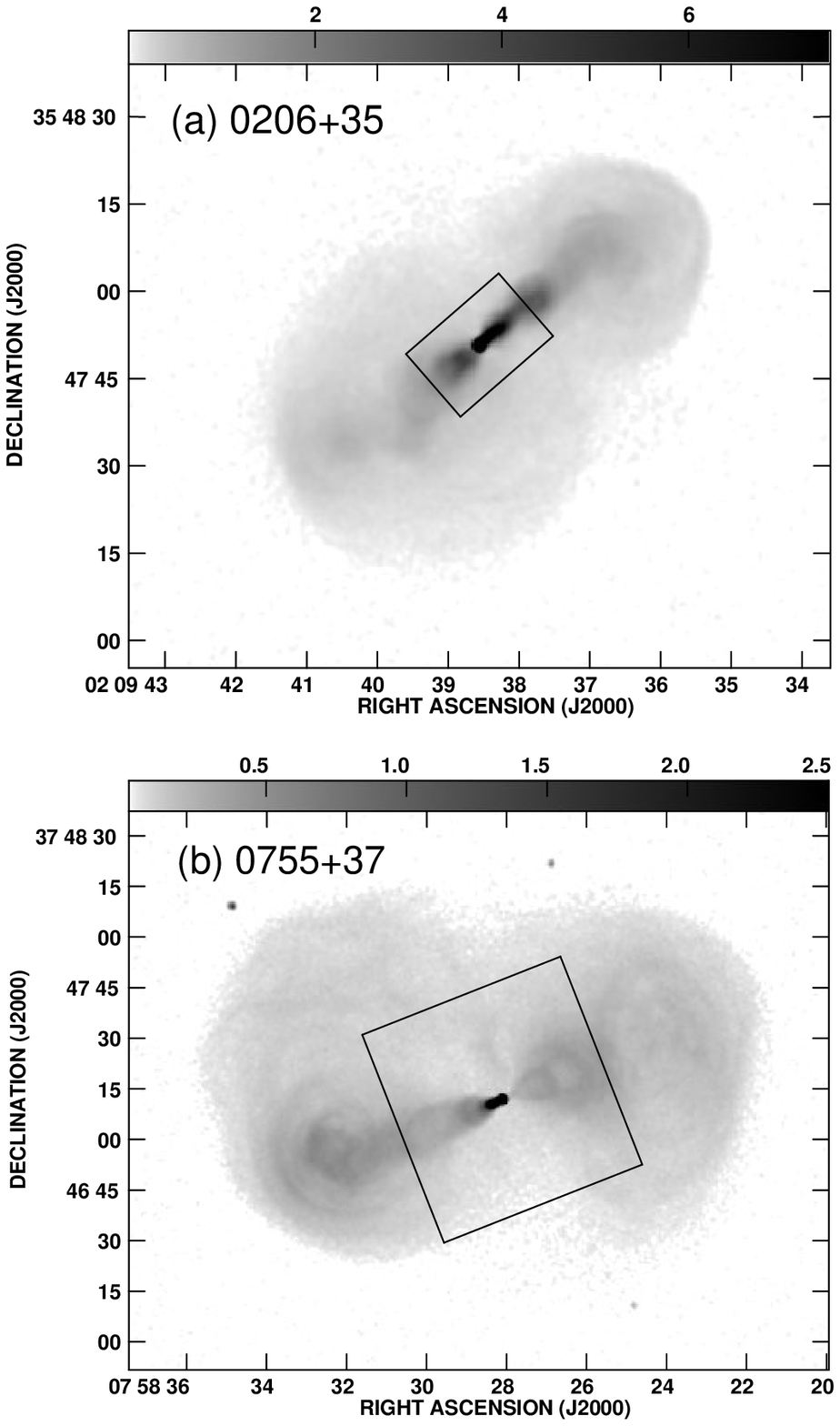}
\caption{Grey-scale images of the sources \citep{LGBPB}. The boxes mark the
  areas shown in later plots and the grey-scale ranges, in mJy\,beam$^{-1}$, are indicated
  by the labelled wedges.  (a) 0206+35 at 4.9\,GHz, 1.2\,arcsec FWHM.  (b)
  0755+37 at 4.9\,GHz, 1.3\,arcsec FWHM.
\label{fig:allsources}
} 
\end{center}
\end{figure}

The galaxy identifications, redshifts and linear scales for the two sources studied
here are given in Table~\ref{tab:sources}. Their radio structures have
been described in detail by \citet{LGBPB}, from which the images
in Fig.~\ref{fig:allsources} are taken.

\begin{center}
\begin{table}
\caption{Names, redshifts, linear scales and associated references for the 
  sources in this paper.\label{tab:sources} }
\begin{tabular}{lllll}
\hline
Name & Galaxy &  Redshift & Scale              & Reference \\
     & name   &           & kpc &            \\
     &        &           & arcsec$^{-1}$ & \\
\hline
0206+35 & UGC\,1651 & 0.03773 & 0.748 & 1 \\
0755+37 & NGC\,2484 & 0.04284 & 0.845 & 2  \\
\hline 
\multicolumn{5}{l}{\scriptsize References: (1) \citet{Miller02}; (2) \citet{Falco99}.}
\end{tabular}
\end{table}
\end{center}

\begin{figure}
\begin{center}
\epsfxsize=8.5cm
\epsffile{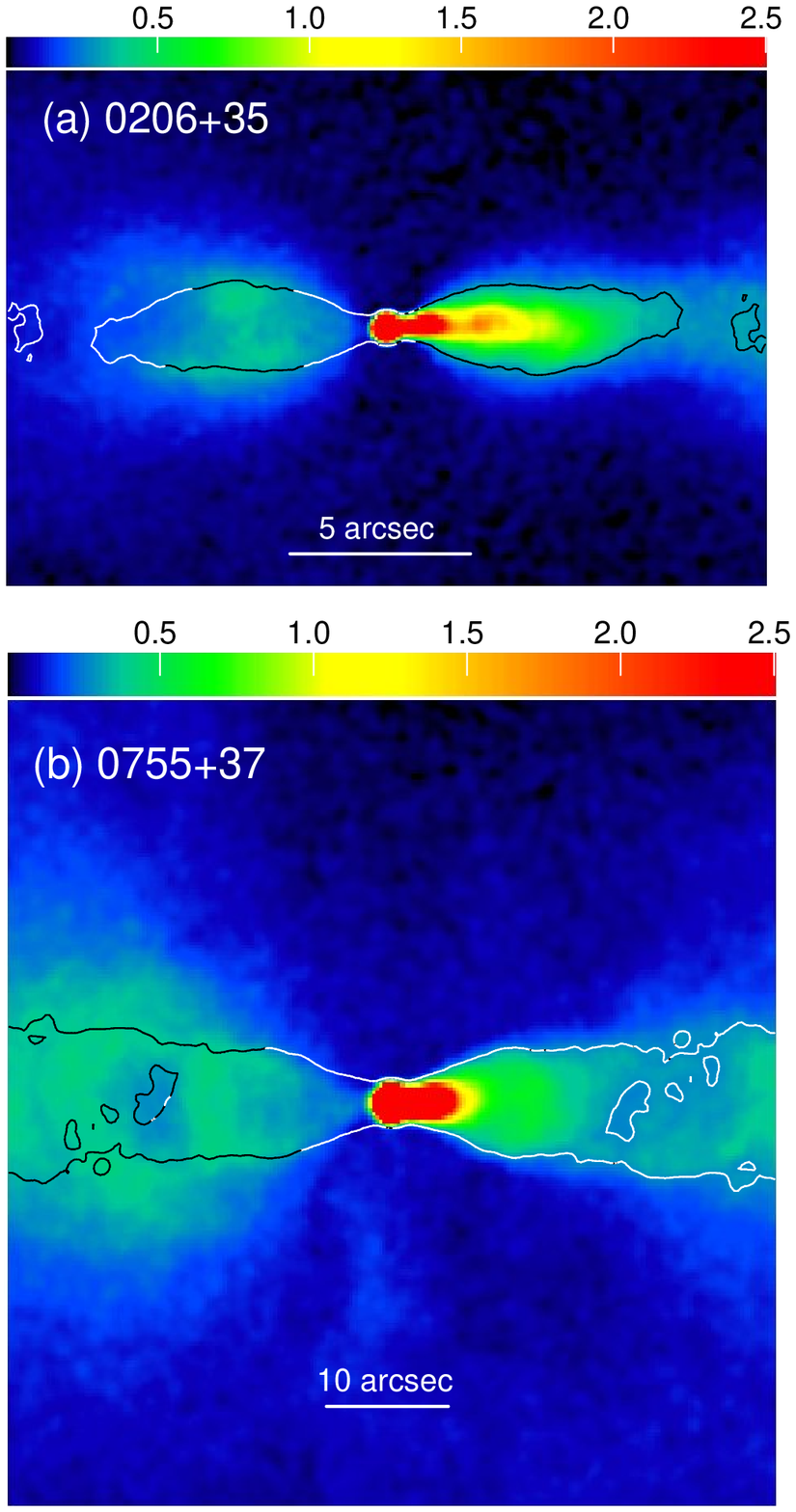}
\caption{False-colour images of total intensity for 0206+35 and 0755+37 over the
  areas outlined in Fig.~\ref{fig:allsources}. On the right-hand side of each
  panel, we have plotted a single contour to outline the brightest emission of
  the main (brighter) jet. On the left-hand side, this contour (rotated through
  180\degr) is plotted on the counter-jet emission. (a) 0206+35 at 0.35-arcsec
  FWHM resolution. (b) 0755+37 at 1.3-arcsec FWHM resolution.
\label{fig:colcont}
}
\end{center}
\end{figure}

\subsection{Images}
\label{images}

\begin{center}
\begin{table}
\caption{Parameters of the sub-images used for modelling and spectral
  analysis. Col. 1: source name; col. 2: observing frequency (an asterisk indicates
  that the image was used for modelling); col
  3: resolution (FWHM); col. 4: rms off-source noise level in $I$; col.5:
  average noise level in $Q$ and $U$; col. 6: sub-image position angle; col. 7:
  sub-image sizes parallel and perpendicular to the jet axis.
\label{tab:images}}
\begin{tabular}{lllrrrr}
\hline
Source & $\nu$ & Res   &\multicolumn{2}{c}{rms} & Rot & Size \\ 
       & GHz & arc- &\multicolumn{2}{c}{$\mu$Jy\,b$^{-1}$}& deg & arcsec$^2$\\
       &  &  sec       & $\sigma_I$ & $\sigma_P$ & &\\
\hline
0206+35 & 1.425 & 1.20 &  19 & $-$   & $-$41.0&$22 \times 20$\\
0206+35 & 4.860 & 1.20 &  12 & $-$   & $-$41.0&$22 \times 20$\\
0206+35 & 4.860 & 0.35*& 7.2 & 7.1   & $-$41.0&$22 \times 20$\\
0755+37 & 1.425 & 1.30 & 20  & $-$   &   158.5&$66 \times 66$\\
0755+37 & 4.860 & 1.30*& 7.8 & 7.9   &   158.5&$66 \times 66$\\
0755+37 & 4.860 & 0.40*& 8.0 & 7.1   &   158.5&$20 \times 16$\\
\hline
\end{tabular}
\end{table}
\end{center}

Table~\ref{tab:images} summarizes the relevant parameters of the
high-resolution sub-images which we model or use for spectral analysis
in this paper (details of the observations and data reduction are
given by \citealt{LGBPB}).  The ${\bf E}$-vector position angles of
linear polarization at 4.860\,GHz have been corrected for Faraday
rotation using multifrequency imaging \citep{Bands,0755RM,LGBPB} and
residual depolarization is predicted to be negligible at this
frequency. The areas plotted in later figures are outlined on
Fig.~\ref{fig:allsources}.

Fig.~\ref{fig:colcont} shows rotated sub-images. On the right-hand side of each
panel, we have drawn a single contour to outline the brightest part of the main
jet. On the left-hand side, this contour is rotated through 180\degr and plotted
on the counter-jet emission. This diagram emphasizes the points made earlier
that the minima in the counter-jet emission have roughly the same widths as the
main jets and that the main jets are in turn surrounded by fainter emission.

In order to model jets that appear superimposed on lobes, we must
try to separate the two emission components in all Stokes parameters.  There is no unique
way to do this when their spectra and intensities vary
independently across the field of view.  Any approach to isolating jet emission
in a lobed FR\,I source therefore entails some simplifying assumption about the
variations in intensity $I$ or in spectral index\footnote{We define spectral
  index $\alpha$ in the sense $I(\nu) \propto \nu^{-\alpha}$.}  $\alpha$ of the
lobes or jets over the region to be modelled.  We have attempted to separate the
jets and lobes for these sources in a way that optimizes the resolution and
signal-to-noise of the jet emission while letting us check for systematic errors
resulting from the assumptions made while doing the lobe-jet separation, as
follows.

One approach to separating jet and lobe emission observed at two
frequencies is based on their systematic {\it spectral} differences: the 
jets have characteristic spectral indices close to $\alpha = 0.55$, whereas the lobes
have $\alpha \ga 0.8$ near the centres of the sources \citep{LGBPB}.  If the 
spectral index of the lobe emission close to the jet is reasonably constant, 
we can use a variant of the `spectral tomography' method \citep{K-SR,KSetal,LCCB06}
by assuming that what is observed can be described as the sum of two components: a
jet and a lobe with constant spectral indices $\alpha_{\rm j}$ and $\alpha_{\rm
  l}$, respectively.  The brightnesses observed at a given point at two
frequencies $\nu_0$ and $\nu_1$ are then:
\begin{eqnarray*}
I(\nu_0) & = & B_{\rm j}\nu_0^{-\alpha_{\rm j}} +  B_{\rm l}\nu_0^{-\alpha_{\rm
    l}}\\  
I(\nu_1) & = & B_{\rm j}\nu_1^{-\alpha_{\rm j}} +  B_{\rm l}\nu_1^{-\alpha_{\rm
    l}}\\  
\end{eqnarray*}
We can scale and subtract the two brightness distributions to estimate the jet brightness 
at the modelling frequency $\nu_0$:
\begin{eqnarray*}
B_{\rm j}\nu_0^{-\alpha_{\rm j}} & = & \frac{\nu_0^{\alpha_{\rm l}}I(\nu_0) -
  \nu_1^{\alpha_{\rm l}}I(\nu_1)}{\nu_0^{\alpha_{\rm l}}-\nu_1^{\alpha_{\rm
      l}}(\nu_0/\nu_1)^{\alpha_{\rm j}}}
  \\ 
\end{eqnarray*}
Once we know $\alpha_{\rm l}$, the method can also
be applied to Stokes $Q$ and $U$ provided that we correct the images at both
frequencies for Faraday rotation before subtraction, and that depolarization is
negligible (as is the case for these sources). Note that the spectral index of
the jets, $\alpha_{\rm j}$, must be both constant and known in order to scale the
result correctly.

In practice, we estimated the lobe spectral index for $\nu_0 = 4.860$\,GHz and
$\nu_1 = 1.425$\,GHz by performing the subtraction
for various trial values of $\alpha_{\rm l}$ and selecting that which minimized
the residual lobe emission in jet-free regions.
 
Spectral subtraction can remove even rather complicated
lobe emission if the spectral index is constant, but it has two serious
flaws for our purposes: (a) the signal-to-noise ratio of the corrected image is
lower than that of the deep high-frequency image alone and (b) our
highest-resolution data for 0206+35 and 0755+37 are only at one frequency.  

The alternative of {\it spatial} subtraction assumes that the lobe {\sl intensity} 
varies only slowly across the jet. This approach can be best applied at high 
angular resolution where the lobe brightness is low and the spatial variation of jet 
emission is clearest. 
To separate the two types of emission spatially in $I$, $Q$ and $U$, we 
define two background regions parallel to the jet axis and just outside the
maximum transverse extent of the jet as estimated from spectral-index images,
i.e.\ using both the intensity and spectral properties of the jet emission to
guide our choice of the background regions.  We then smooth the background 
brightness distributions parallel to the jet axis with a boxcar function to
improve their signal-to-noise ratio and interpolate linearly between
them under the jet.\footnote{Higher-order interpolation works poorly for these
  brightness distributions.}   We refer to this approach as generating `interpolated
images'.

For 0206+35 and 0755+37 we first used spectral
subtraction to verify the total extent of the jet emission and to set
appropriate reference regions for interpolation, then constructed interpolated images for the
final modelling.

\begin{table}
\caption{Interpolation parameters for lobe subtraction. Col. 1: source name;
  col 2: resolution (FWHM); col. 3: background region distances from jet axis; col. 4: width of boxcar
  smoothing function parallel to the axis.\label{tab:interp}}
\begin{tabular}{llrr}
\hline
Source & FWHM & Background & Smooth  \\
       &  (arcsec)  &  (arcsec)  & (arcsec)\\
\hline
0206+35& 0.35 & 9 -- 10 & 1.0 \\
0755+37&1.30  & 30 -- 45 & 3.0 \\
\hline
\end{tabular}
\end{table}

\begin{figure}
\begin{center}
\epsfxsize=8.5cm
\epsffile{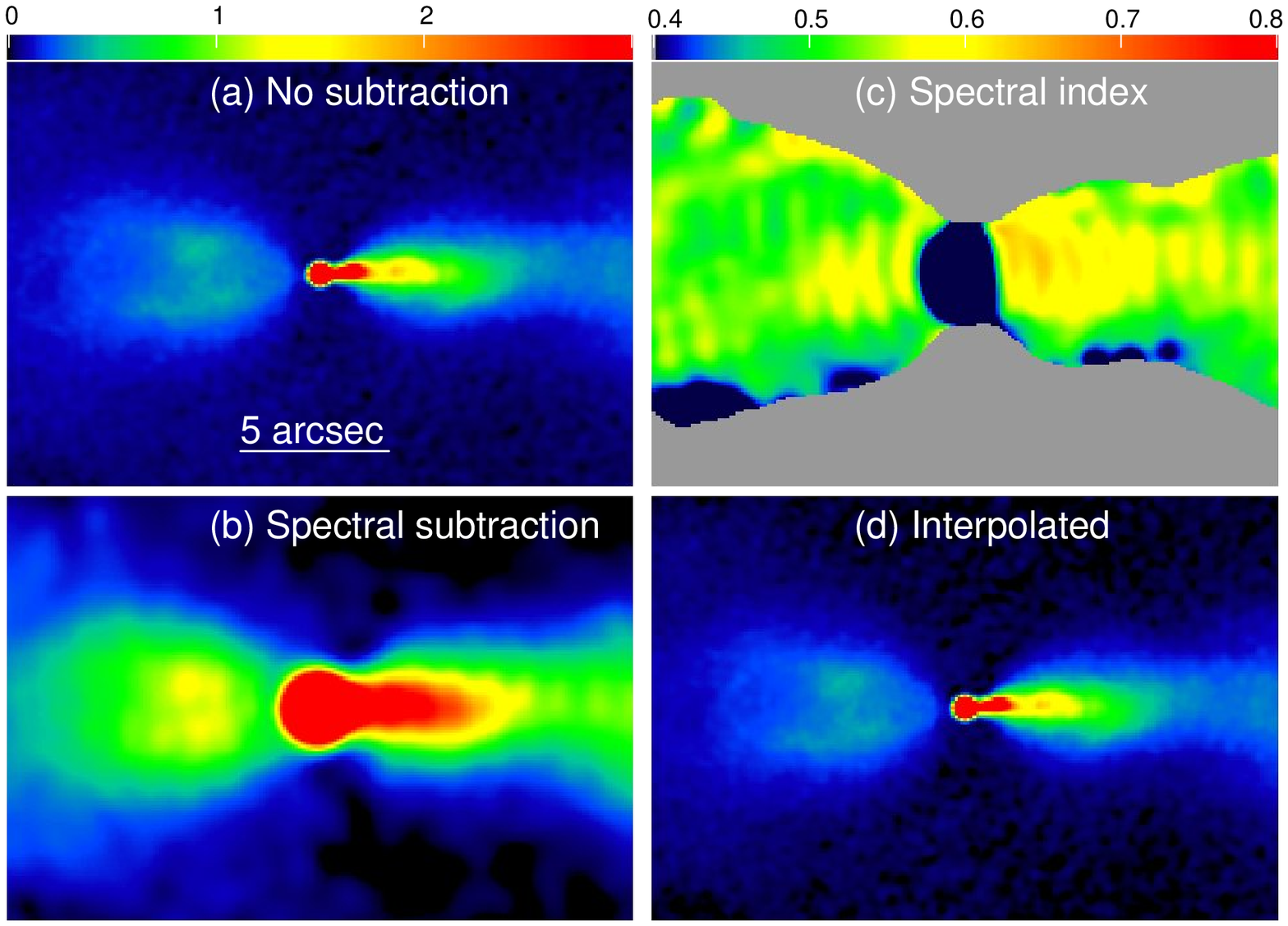}
\caption{False-colour images showing the results of lobe subtraction for
  0206+35. The $I$ intensity colour range (0 -- 3\,mJy\,beam$^{-1}$) is the same for 
  panels (a) and (d). (a) No subtraction at 0.35-arcsec resolution. (b) Subtraction at 1.2-arcsec resolution
  assuming a constant spectral index for the lobe. (c) Spectral index distribution 
  over the jet and counter-jet at 1.2-arcsec resolution after interpolated
  subtraction,  
  blanked where $\sigma_{\alpha} > 0.03$
  (the colour range for spectral index is shown by the labelled wedge).
  (d) Subtraction by linear interpolation between background strips parallel to
  the jet axis. The resolution is 0.35\,arcsec.
\label{fig:0206sub}
}
\end{center}
\end{figure}

\begin{figure}
\begin{center}
\epsfxsize=8.5cm
\epsffile{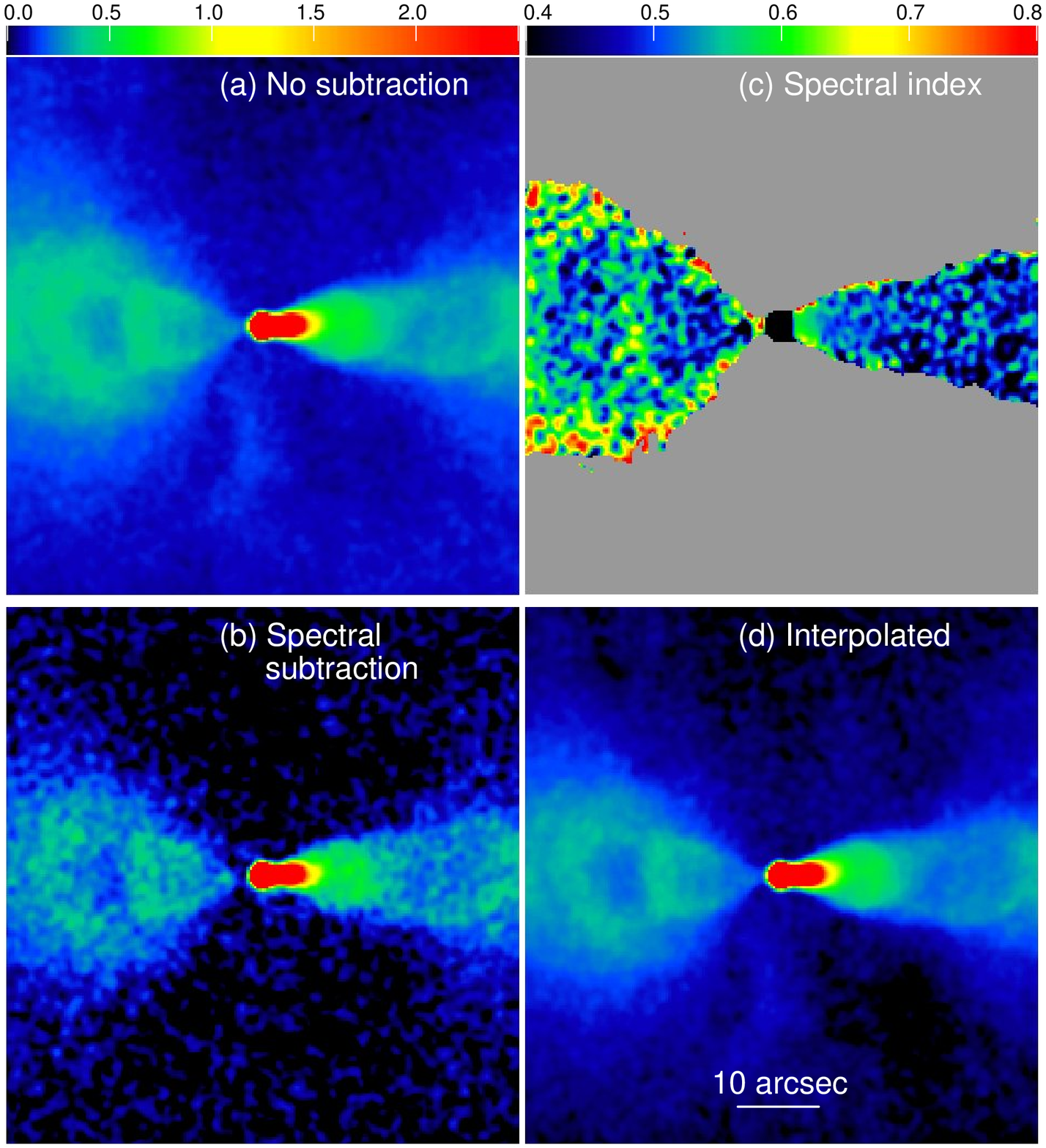}
\caption{False-colour images showing results of lobe subtraction in 0755+37 at
  1.3-arcsec resolution. 
  The $I$ intensity colour range (0 -- 2.5\,mJy\,beam$^{-1}$) is the same for 
  panels (a), (b) and (d). (a) No subtraction. (b) Subtraction assuming a constant
  spectral index for the lobe, as described in the text.  (c) Spectral index distribution 
  over the jet and counter-jet after interpolated lobe subtraction, 
  blanked where $\sigma_{\alpha} > 0.1$ (the colour range for spectral index is
  indicated by the labelled wedge).
  (d) Subtraction by
  linear interpolation between background strips parallel to the jet axis. 
\label{fig:0755sub}
} 
\end{center}
\end{figure}

In Figs~\ref{fig:0206sub} and \ref{fig:0755sub}, we show the results of both
subtraction methods for the two sources. Figs~\ref{fig:0206sub}(a) and
\ref{fig:0755sub}(a) show the images at the resolution used for modelling before
subtraction.  We found best-fitting lobe spectral indices between 1.425 and
4.860\,GHz of 0.90 and 0.81 for
0206+35 and 0755+37, respectively.  In Fig.~\ref{fig:0206sub}(b), we show the spectral
subtraction for 0206+35 at lower resolution. Although not useful for modelling,
this image outlines the total extent of the flatter-spectrum emission associated
with the jets.  The spectral subtraction for 0755+37 at the lower of the two
resolutions used for modelling, shown in Fig.~\ref{fig:0755sub}(b), has little
trace of residual lobe emission but low signal-to-noise.  

Guided by the spectral subtraction, we set the interpolation parameters as in
Table~\ref{tab:interp} and computed interpolated images at 1.425 and
4.860\,GHz, from which we in turn derived the spectral-index images shown in
Figs~\ref{fig:0206sub}(c) and \ref{fig:0755sub}(c). These are blanked on the
error in spectral index, as noted in the captions. We then estimated integrated
spectral indices for the jets by summing the interpolated $I$ images over all
pixels which are unblanked on the spectral-index images, excluding the cores. We
found $\langle \alpha_{\rm j}\rangle = 0.55$ for 0206+35 and 0.53 for 0755+37.  We used these
values to scale the spectral subtractions. Variations across
the modelled regions are small, with $0.50 \leq \alpha_{\rm j} \leq 0.62$ in both
sources. Finally, we show the 4.860-GHz
interpolated images at the resolutions used for modelling in
Figs~\ref{fig:0206sub}(d) and \ref{fig:0755sub}(d).

We are confident that the interpolated images represent the jet emission
accurately in both sources.  The lobe emission in 0206+35 is quite faint at
0.35-arcsec FWHM resolution, and after subtraction, the area around the jets appears
devoid of residual emission in all Stokes parameters (e.g.\
Fig.~\ref{fig:0206sub}d). The lower-resolution (1.3 arcsec FWHM) image of
0755+37 proved to be more of a challenge, because the lobe emission is bright
and irregular (Fig.~\ref{fig:0755sub}a).  The spectral subtraction gave a clean
image of the jet with negligible background emission, but amplified noise
(Fig.~\ref{fig:0755sub}b).    In contrast, interpolation (Fig.~\ref{fig:0755sub}d)
failed to remove the small-scale lobe emission accurately but retained the full
signal-to-noise ratio of our high-frequency images. Comparison of the two
corrected $I$ images showed that they are accurately consistent wherever $I >
100$\,$\mu$Jy\,beam$^{-1}$.  We therefore used the interpolated images for
modelling (in which the faint residual lobe emission has low weight). Modelling the spectrally-subtracted image
(and its counterparts in $Q$ and $U$) gave consistent but
less well constrained results.  In the intensity and polarization profiles
plotted below, we compare the results from both subtraction methods.

At 0.4-arcsec FWHM resolution, used for modelling the inner jets of 0755+37, the lobe
brightness is negligible and we did not attempt to subtract it.

\section{Model fits}
\label{fits}

\subsection{Assumptions}
\label{model-assumptions}

To model the jet emission, we make the following assumptions.
\begin{enumerate}
\item The jets are intrinsically symmetrical, axisymmetric and 
  antiparallel. They can be treated, on average, as laminar, stationary flows.
\item The radio emission is from relativistic particles with a power-law energy
  spectrum $n(E) = n_0 E^{-(2\alpha+1)}$ ($\alpha$ is the spectral index). We
    use the integrated values for the modelled regions after lobe subtraction: $\langle \alpha\rangle =
    0.55$ for 0206+35 and 0.53 for 0755+37.
    The corresponding maximum degree of polarization is  
    $p_0 = (3\alpha+3)/(3\alpha+5) = 0.70$ in both cases and the variations of
    spectral index across the modelled regions are small enough to be ignored (Section~\ref{images}).
\item The magnetic field is tangled on small scales, but anisotropic.
\item The effects of Faraday rotation on the observed emission are 
 corrected completely. This is an extremely good approximation for 0206+35 and
 0755+37 \citep{Bands,0755RM,LGBPB}.
\end{enumerate}

\subsection{Outline of method}
\label{model-outline}

For a symmetrical, outflowing jet with velocity $v = \beta c$, emitting isotropically in
the rest frame and inclined by an angle $\theta$ to the line of sight, a measurement of the observed jet/counter-jet intensity ratio
\begin{eqnarray*} 
I_{\rm j}/I_{\rm cj} &=& [(1+\beta\cos\theta)/(1-\beta\cos\theta)]^{2+\alpha} \\
\end{eqnarray*}
does not allow us to determine the velocity and inclination separately.  The key
to our method is the use of linear polarization to break this degeneracy.  The
relation between the angles to the line of sight in the rest frame of the outflow,
$\theta^\prime$ and in the observed frame, $\theta$, is:
\begin{eqnarray*} 
\sin\theta^\prime_{\rm j} & = & [\Gamma(1-\beta\cos\theta)]^{-1}\sin\theta
\makebox{~~~~~(main jet)} \\
\sin\theta^\prime_{\rm cj} & = & [\Gamma(1+\beta\cos\theta)]^{-1}\sin\theta
\makebox{~~(counter-jet)} \\
\end{eqnarray*}
The emission in all three Stokes parameters depends on $\theta^\prime$, since the magnetic
field is in general anisotropic. If the flow is significantly relativistic, we
effectively observe the two jets at different values of $\theta^\prime$ and can
use the differences in polarization for the approaching and receding jets as an
additional constraint to separate $\beta$ and $\theta$. For backflow, the
argument is identical with the roles of jet and counter-jet interchanged.

The principal steps in our method \citep{LB02a,CL,CLBC,LCBH06} are as follows.
\begin{enumerate}
\item Build a parameterized model of the geometry, the velocity field and the
  variations of  
  emissivity ($\propto n_0
  B^{1+\alpha}$) and magnetic-field anisotropy in the rest frame of the emitting
  plasma.
\item Calculate the observed-frame emission in $I$, $Q$ and $U$, taking account
  of relativistic aberration and anisotropic emission in the rest frame.
\item Integrate along the line of sight, normalize to the measured total flux
  density and convolve with the observing beam.
\item Calculate and sum $\chi^2$ over the $I$, $Q$ and $U$ images. This is our
  measure of goodness of fit.
\item Optimize the parameters using the downhill simplex method of Nelder \& Mead \citep{NR}.
\end{enumerate}
We explored a wide range of starting simplexes in order
to be sure of locating the global minimum in $\chi^2$.

\subsection{Fitting functions}
\label{fit-funcs}

The parameterized model that we fit to the VLA observations is a simplified
version of 
those in our previous work \citep{LB02a,CL,CLBC,LCBH06}, with the addition of a few extra terms to describe
the backflow.   The functional forms are  
given explicitly in Table~\ref{tab:functions}.  A critical 
discussion of fitting functions will be given elsewhere (Laing \& Bridle,
in preparation).

\begin{table*}
\caption{Coordinate definitions and functional forms for geometry, velocity, proper emissivity and
  magnetic-field ordering.\label{tab:functions}}
\begin{tabular}{llll}
\hline
&&&\\
Description & Quantity & Functional form & Distance range\\
&&&\\
\hline
&&&\\
Distance coordinate &
$r$ & $\frac{zr_0}{(r_0 + A)\cos\xi -A}$&$r \leq r_0$ \\
(outflow and backflow)&& $\frac{z+A}{\cos\xi} - A$& $r \geq r_0$\\
    && $A = x_0/\sin\xi_0-r_0 = x_{\rm b}/\sin\xi_{\rm b}-r_0$&\\
&&&\\
Outflow streamline index & $s$ & by continuity  & $r \leq r_0$ \\ 
&& $\xi/\xi_0$ & $r \geq r_0$ \\
&&&\\
Outflow radius & $x(z,s)$ & $a_2(s)z^2 + a_3(s)z^3$ & $r \leq r_0$ \\
         && $(z-r_0+x_0/\sin\xi_0)\tan(\xi_0s)$& $r \geq r_0$ \\
&&&\\
Outflow velocity &$\beta(r,s)$ & $\beta_1 \exp(s^2\ln v_1)$ & $r \leq r_{v_1}$ \\
             && $\beta_1 \exp(s^2\ln v_1)\left (\frac{r_{v0} - r}{r_{v0} -r_{v1}}
             \right ) + \beta_0 \exp(s^2\ln v_0)\left (\frac{r - r_{v1}}{r_{v0} -r_{v1}}
             \right )$ & $r_{v1} \leq r \leq r_{v0}$ \\
             && $\beta_0 \exp(s^2\ln v_0)$ & $r \geq r_{v_0}$ \\
&&&\\
Outflow proper emissivity &$\epsilon(r,s)$&$g_1 r^{-E_{\rm in}}\exp(\ln e_1 s^2)$&$r \leq r_{e1}$\\
               &&$~~r^{-E_{\rm mid}}\exp \left[\ln \left
                   (\frac{e_1(r_{e0}-r)+e_0(r-r_{e1})}{r_{e0}-r_{e1}} \right )
                 s^2\right ]$&$r_{e1} \leq r \leq r_{e0}$\\ 
               &&$g_0 r^{-E_{\rm out}}\exp(\ln e_0 s^2)$&$r \geq r_{e0}$\\
&&&\\
Outflow $\langle B_r^2/B_t^2\rangle^{1/2}$ &
$j(r)$&$j_1 $&$r \leq r_{B1}$\\ 
      &&$\frac{j_1(r_{B0}-r)+j_0(r-r_{B1})}{r_{B0}-r_{B1}}$&$r_{B1} \leq r \leq r_{B0}$\\
      &&$j_0 $&$r \geq r_{B0}$\\ 
Outflow $\langle B_l^2/B_t^2\rangle^{1/2}$ &
$k(r)$&$k_1 $&$r \leq r_{B1}$\\ 
      &&$\frac{k_1(r_{B0}-r)+k_0(r-r_{B1})}{r_{B0}-r_{B1}}$&$r_{B1} \leq r \leq r_{B0}$ \\
      &&$k_0 $&$r \geq r_{B0}$\\ 
&&&\\
Backflow streamline index& $t$ & by continuity  & $r \leq r_0$ \\ 
&& $(\xi-\xi_0)/(\xi_{\rm b}-\xi_0)$& $r
\geq r_0$\\
&&&\\
Backflow radius & $x(z,t)$ & $a_2(t)z^2 + a_3(t)z^3$ & $r \leq r_0$ \\
         && $(z-r_0+x_0/\sin\xi_0)\tan[\xi_0+(\xi_{\rm b}-\xi_0)t]$& $r \geq r_0$ \\
&&&\\
Backflow velocity &
$\beta(t)$ & $\beta_{\rm b, in} + t(\beta_{\rm b, out}-\beta_{\rm b, in})$\\
&&&\\
Backflow proper emissivity &
$\epsilon(r,t)$& 0              & $r < r_b$\\
             && $n_{\rm b}(r/r_0)^{-E_{\rm b}}\exp(\ln e_{\rm b}t^2)$& $r \geq r_b$ \\
&&&\\
Backflow $\langle B_r^2/B_t^2\rangle^{1/2}$&
$j$& $j_{\rm b}$ &\\
Backflow $\langle B_l^2/B_t^2\rangle^{1/2}$& $k$& $k_{\rm b}$ &\\
&&&\\
\hline
\end{tabular}
\end{table*}

\begin{figure}
\begin{center}
\epsfxsize=6cm
\epsffile{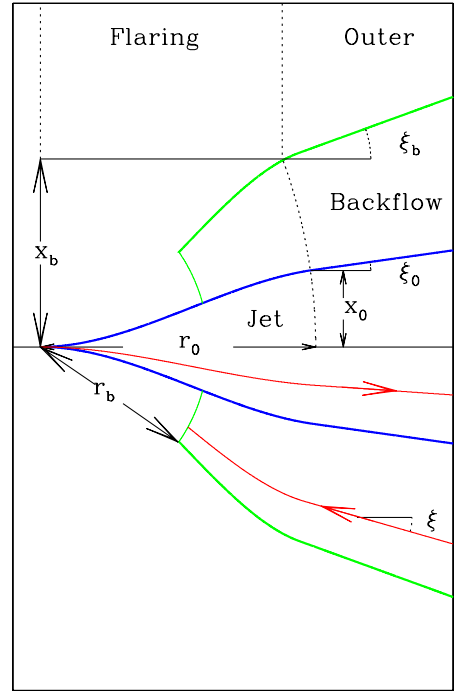}
\caption{Sketch of the assumed geometry. The blue and green curves show the
  outer boundaries of the outflowing jet and backflow emission, respectively,
  Representative streamlines in the two parts of the flow are shown in red. The
  fiducial distances and angles are defined in Section~\ref{geom-functions}.
\label{fig:sketch}
}
\end{center}
\end{figure}
\subsubsection{Geometry}
\label{geom-functions}

We use coordinates $(z,x)$ in a plane containing the jet axis, with $z$ measured
along the axis and $x$ perpendicular to it.  The jet is divided into a {\em
flaring region}, where the flow first expands and then recollimates, and a
conical {\em outer region}, as sketched in Fig.~\ref{fig:sketch}. The edge of
the outflow is fully defined by the distance of the transition between the two
regions measured along the axis, $r_0$, the radius, $x_0$, and the opening angle
of expansion in the outer region, $\xi_0$. Individual streamlines in the outer
region are straight, so we can define a streamline index $s = \xi/\xi_0$, where $\xi$
is the angle between the streamline and the axis.  $s$ ranges from 0 on-axis to 1
at the edge of the outflow. The two coefficients $a_2(s)$ and $a_3(s)$ of the
cubic expression for the streamline radius in the flaring region (Table~\ref{tab:functions}) are
defined by the conditions that the streamline radius $x(z)$ and its first
derivative $x^\prime(z)$ are continuous at the flaring-outer region boundary.
We also define a distance coordinate $r$ which is continuous along a given
streamline from 0 at the nucleus to $r_0$ at the flaring-outer region boundary
and which thereafter increases as the distance from the boundary surface. The
functional forms for $r$ in the two regions are given in terms of $z$ and
$s$ in Table~\ref{tab:functions}

For simplicity, the backflow is assumed to follow the same streamline family as
the jet, extended away from the axis.  The edge of the backflow in the outer
region is defined by the radius $x_{\rm b}$ at the region boundary and the
opening angle $\xi_{\rm b}$.  These are not independent: $x_{\rm
  b}/x_0 = \sin\xi_{\rm b}/\sin\xi_0$. The backflow streamline index $t$ ranges from 0
at the backflow/outflow interface to 1 at the edge of the outflow.  The backflow
streamline radii have the same functional form as their outflow equivalents with
the coefficients $a_2(t)$ and $a_3(t)$ again defined by continuity at the region
boundary.

The assumed backflow geometry is ad hoc, but gives a reasonable match 
to the observed extent of the emission.

\subsubsection{Velocity}
\label{vel-functions}

The on-axis velocity profile in the outflow is divided into three parts: (a)
constant with a high velocity close to the nucleus; (b) a linear
decrease and (c) constant with a low 
velocity at large distances.  The velocity along any off-axis streamline is calculated 
using the same expressions but with truncated Gaussian transverse profiles.  The velocity
profiles, given explicitly in Table~\ref{tab:functions}, depend on two
transition distances, $r_{v1}$ and $r_{v0}$, the on-axis velocities $\beta_1$
and $\beta_0$ and the fractional edge velocities $v_1$ and $v_0$ (which are
required to be $\leq 1$).

We experimented with several functional forms for the backflow velocity. The
most satisfactory has no dependence on $r$, but varies linearly with streamline
index from $\beta_{\rm b, in}$ at the interface with the outflow to $\beta_{\rm b, out}$ at
the outer edge of the backflow. 

\subsubsection{Emissivity}
\label{em-functions}

We write the proper emissivity as $\epsilon f$, where $\epsilon$ is
the emissivity in Stokes $I$ for a magnetic field perpendicular to the line of
sight and $f$ depends on the field geometry (defined in
Section~\ref{mag-functions}, below). $\epsilon$, to which we refer loosely as
`the emissivity', is a function only of the total rms magnetic-field strength
and the normalizing constant of the radiating electron energy distribution.

The on-axis emissivity profile in the outflow is also divided into three
regions, each with a power-law profile. The profile is allowed to be
discontinuous at each of the region boundaries. Off-axis, the profile is
multiplied by a truncated Gaussian function of the streamline index, with
values at the jet edge which are constants in the inner and outer emissivity regions and
vary linearly between them. The free parameters for the emissivity profiles are
transition distances, $r_{e0}$ and $r_{e1}$, power-law indices $E_{\rm in}$,
$E_{\rm mid}$ and $E_{\rm out}$, $g_1$ and $g_0$, which measure the
discontinuities at the region boundaries and edge emissivities $e_1$
and $e_0$. Note that $e_1$ and $e_0$ may be $>1$ (in which case the jet is
limb-brightened), $= 1$ (uniformly filled) or $<1$ (centre-brightened).

The backflow emissivity is assumed to be zero within a given distance and to have a
power-law dependence on $r$ with a single index and a truncated Gaussian
dependence on streamline index $t$ elsewhere. The fitted parameters are the
index $E_{\rm b}$, the fractional edge emissivity $e_{\rm b}$, the inner
distance $r_{\rm b}$ and the emissivity ratio between outflow and backflow at
the boundary between the flaring and outer regions, $n_b$.

\subsubsection{Magnetic-field structure}
\label{mag-functions}

We define the rms components of the magnetic field to be $\langle
B_l^2\rangle^{1/2}$ (longitudinal, parallel to a streamline), $\langle
B_r^2\rangle^{1/2}$(radial, orthogonal to the streamline and outwards from the
jet axis) and $\langle B_t^2\rangle^{1/2}$ (toroidal, orthogonal to the
streamline in an azimuthal direction).  The rms total field strength is $B =
\langle B_l^2 + B_r^2 + B_t^2\rangle^{1/2}$ The magnetic-field structure is
parameterized by the ratio of rms radial/toroidal field, $j = \langle
B_r^2\rangle^{1/2}/\langle B_t^2\rangle^{1/2}$ and the longitudinal/toroidal
ratio $k =\langle B_l^2\rangle^{1/2}/\langle B_t^2\rangle^{1/2}$. For the
outflow models in the present paper, these depend only on $r$, being constant
close to and far from from the nucleus and varying linearly at intermediate
distances. The free parameters are the fiducial distances $r_{B1}$ and $r_{B0}$
and the field ratios at these distances, $j_1$, $j_0$, $k_1$ and $k_0$.

For the backflow, we assume constant field ratios $j_{\rm b}$ and $k_{\rm
  b}$. 

\subsection{Modelling of individual sources}
\label{model-specifics}

We estimated the noise levels for each resolution and Stokes parameter based on
the deviations of the brightness distributions from those expected for
axisymmetry, as follows.
\begin{enumerate}
\item Calculate Stokes parameters $Q$ and $U$ in a coordinate system with
  position angle 0 along the jet axis.
\item For $I$ and $Q$, take the noise level to be $1/\sqrt{2}$ times the rms
  difference  
between the image and a copy of itself
reflected across the jet axis. 
\item For $U$, take the sum rather than the difference.
\end{enumerate}
These values can be substantially larger than the
off-source rms, but include the effects of small-scale structure (which we do
not attempt to model) and deconvolution errors. 

For 0206+35, we fit to images at the highest available resolution, 0.35\,arcsec
FWHM, using different noise levels for the high-brightness emission close to the
nucleus and the fainter regions farther out. For 0755+37, we fit to 0.4-arcsec
FWHM images of the bright inner jets and 1.3-arcsec images elsewhere.  Small
regions around the cores were excluded from the fits, since we model only
optically-thin emission. The model images given below include point sources with
the appropriate observed flux densities at the locations of the cores. 

The values of $\chi^2$ summed over all Stokes parameters and resolutions were
8012 over 6696 independent points for 0206+35 and 8022 over 5816 points for
0755+37.

\begin{figure*}
\begin{center}
\epsfxsize=13cm
\epsffile{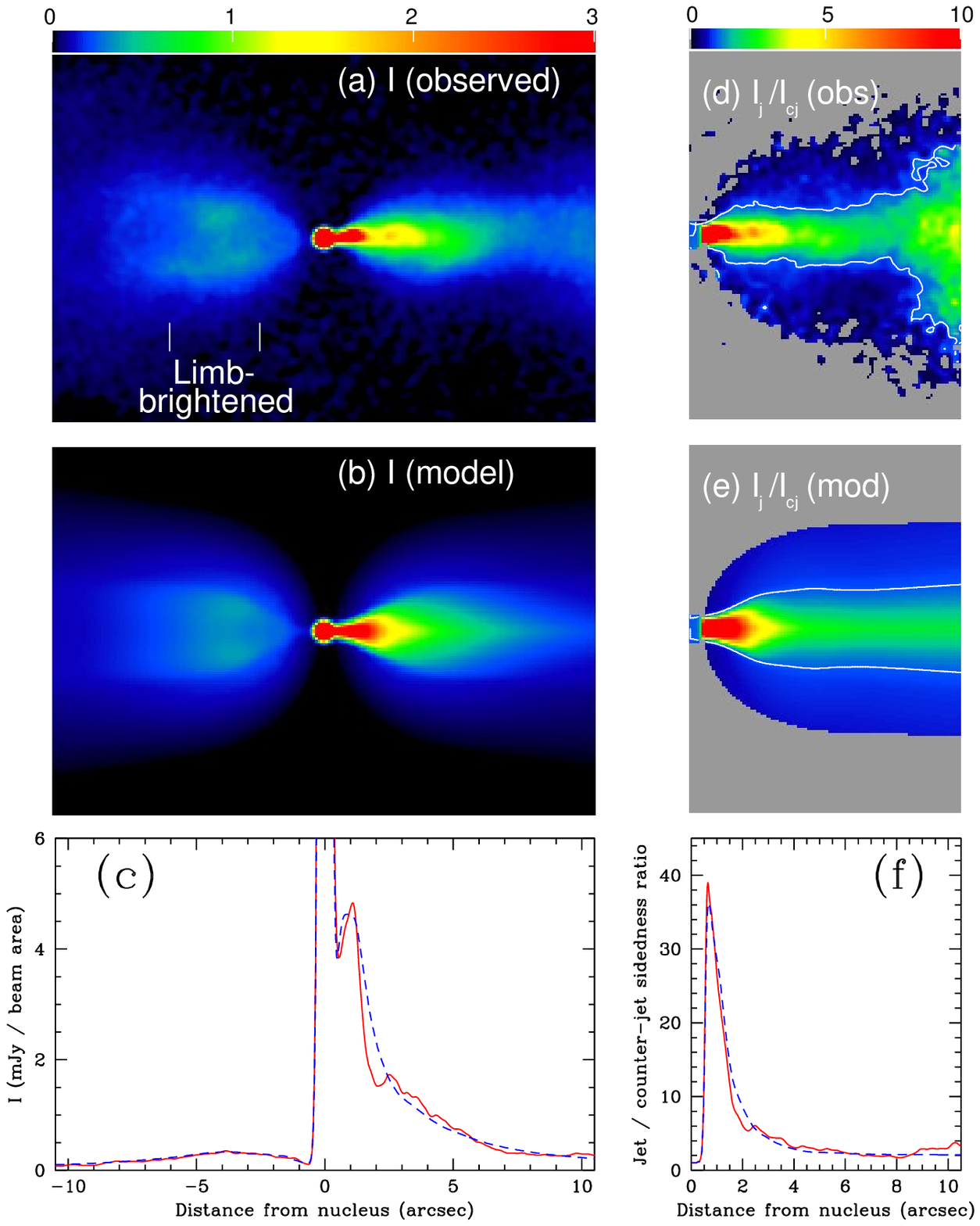}
\caption{Comparison between the observed and modelled total intensities $I$ and
  sidedness ratios $I_{\rm j}/I_{\rm cj}$ for 0206+35. (a) observed and (b)
  model false-colour images of $I$. (c) profiles of observed (full/red) and
  model (dashed/blue) $I$ along the axis of the jet. (d) and (e) images of
  $I_{\rm j}/I_{\rm cj}$. The white contours represent $I_{\rm j}/I_{\rm cj} =
  1$: outside the contours, $I_{\rm j}/I_{\rm cj} < 1$. (f) profiles of observed
  (full/red) and model (dashed/blue) $I_{\rm j}/I_{\rm cj}$ along the jet axis.
\label{fig:0206comp}
} 
\end{center}
\end{figure*}

\begin{figure*}
\begin{center}
\epsfxsize=15cm
\epsffile{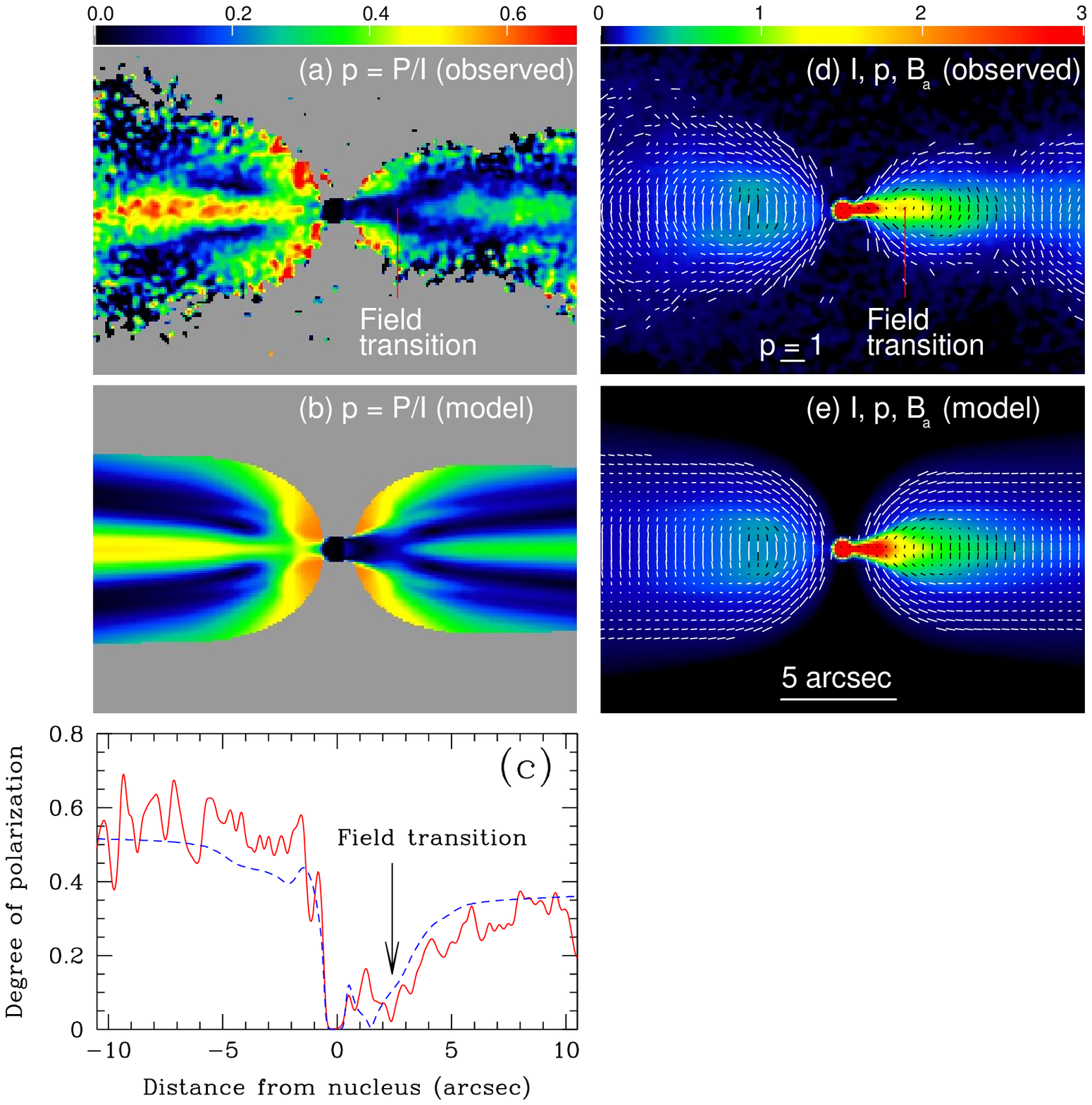}
\caption{Comparison between the observed and modelled linear polarization of
  0206+35. (a) and (b) colour images of degree of polarization $p = P/I$ in the
  range 0 -- 0.7, as indicated by the labelled wedge.  Blanked areas are
  grey. (a) observed; (b) model. (c) profiles of observed (full/red) and model
  (dashed/blue) $p$ along the axis of the jet. (d) and (e) vectors with lengths 
  proportional to $p$ and directions along the apparent magnetic field
  superimposed on colour images of $I$. (d) observed, (e) model. 
\label{fig:0206polcomp}
} 
\end{center}
\end{figure*}

The quoted uncertainties were also derived as in our earlier
work by varying an individual parameter until $\chi^2$ increased by an amount
corresponding to the formal 99 per cent confidence level, leaving the rest of
the model unchanged. These values 
are crude (they neglect coupling between parameters), but in practice give a
good impression of the range of reasonable models. As an additional check, we
also performed a series of optimizations at fixed values of $\theta$ and
tabulate the range over which acceptable solutions could be found.

\section{Model-data comparisons}
\label{compare}

\subsection{General}
\label{gencompare}

\begin{figure*}
\begin{center}
\epsfxsize=14cm
\epsffile{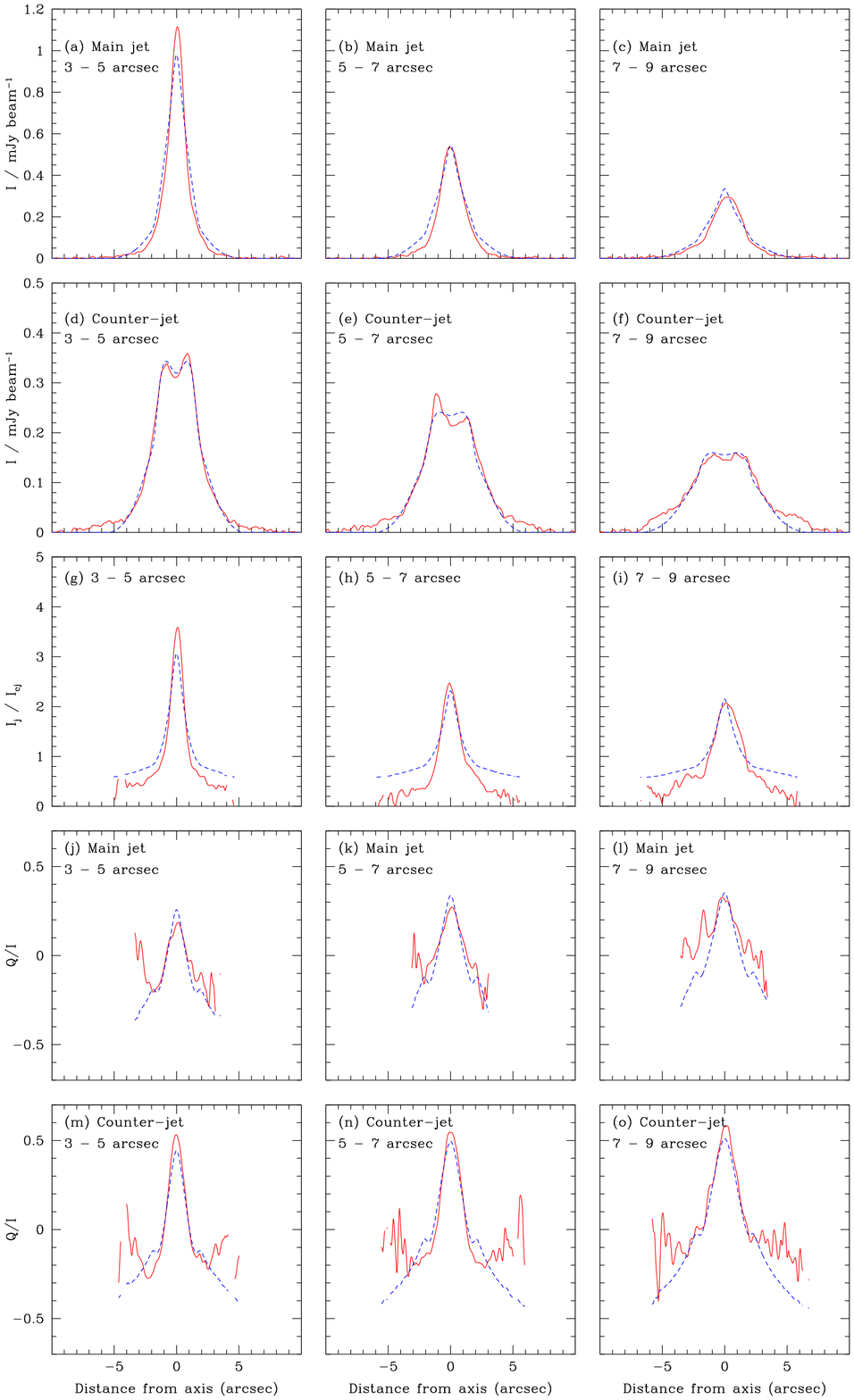}
\caption{Transverse profiles of total intensity, $I$, jet/counter-jet sidedness
  ratio, $I_{\rm j}/I_{\rm cj}$ and $Q /I$ for 0206+35. The data have been
  averaged parallel to the jet axis over three ranges of distance from the
  nucleus: 3 -- 5, 5 -- 7 and 7 -- 9\,arcsec, as indicated in the captions
  (Section~\ref{gencompare}).  Full/red line: observations; dashed/blue line:
  model. $Q/I > 0$ and $Q/I < 0$ correspond to transverse and longitudinal
  apparent field, respectively.
\label{fig:0206trans}
}
\end{center}
\end{figure*}

In Figs~\ref{fig:0206comp} -- \ref{fig:0206trans} and \ref{fig:0755comp} --
\ref{fig:0755trans}, we show various comparisons between the observed and model
images of the two sources.  The images have been rotated by the angles given in
Table~\ref{tab:images} so that the main (approaching) jet points to the right
and the core is either at the centre or the left-hand edge of a plot. The types
of plot are as follows.
\begin{enumerate}
\item False-colour images of total intensity. The angular scale is given on the
  accompanying profiles and the brightness range (in mJy\,beam$^{-1}$) is
  indicated by the labelled wedges.
\item Longitudinal profiles of total intensity.
\item Images of jet/counter-jet sidedness ratio $I_{\rm j}/I_{\rm cj}$ derived
  by dividing the $I$ image by a copy of itself rotated by 180$^\circ$. These
  images are blanked (grey) where $I < 3\sigma_I$ on either side of the core (Table~\ref{tab:images}). The
  contours show $I_{\rm j}/I_{\rm cj} = 1$. Angular scales are again shown on
  the accompanying profiles.
\item Longitudinal profiles of sidedness ratio.
\item Images of degree of polarization, $p = P/I$. These are blanked wherever $I
  < 5\sigma_I$. The angular scale is given on the accompanying profiles and the
  range is indicated by the labelled wedges. $p$ has been corrected for Ricean
  bias \citep{WK}.
\item Profiles of $p$ along the jet axis.
\item Vectors with lengths proportional to $p$ and directions along the
  apparent magnetic field, superposed on false-colour images of $I$. The angular
  and vector scales are indicated by labelled bars.
\item Averaged transverse profiles of total intensity, $I$, sidedness ratio
  $I_{\rm j}/I_{\rm cj}$, and $Q/I$ over selected regions where the brightness
  and polarization distributions vary slowly with distance from the
  nucleus. Stokes $Q$ is defined in a coordinate system with its axis along the
  jet: $Q/I > 0$ for an apparent magnetic field transverse to the axis; $Q/I <
  0$ for a longitudinal field.  In the flaring region, these profiles were
  derived by averaging along radii from the nucleus, in which case they are
  plotted against angle from the jet axis. For the outer region, they are
  averages along lines parallel to the jet axis and are plotted against angular
  distance from the axis.  In order to make a fair comparison, only pixels which
  were not blanked on the observed images were used in the averages.
\end{enumerate}
In general the fits are very good. We examine the correspondence between model
and observed brightness distributions in detail in the next two sub-sections.

\subsection{0206+35}
\label{0206fit}

We show images and longitudinal profiles of total intensity and sidedness ratio
in Fig.~\ref{fig:0206comp} and of degree and direction of polarization in
Fig.~\ref{fig:0206polcomp}. Averaged transverse profiles of $I$, $I_{\rm
  j}/I_{\rm cj}$ and $Q/I$ are given in Fig.~\ref{fig:0206trans}. 

The model accurately reproduces the main features of the brightness and
polarization distributions of 0206+35, including the following.
\begin{enumerate}
\item The main (approaching) jet has a bright base, with a peak at
$\approx$2\,arcsec from the nucleus (Figs~\ref{fig:0206comp}a -- c).
\item The peak sidedness ratio of $I_{\rm j}/I_{\rm cj} \approx 37$ is 
  at a distance of $\approx$0.6\,arcsec from the nucleus (Figs~\ref{fig:0206comp}d -- f), close to the position
  of the flaring point as determined from high-resolution MERLIN observations
  \citep{LGBPB}.
\item At low isophotes, the counter-jet appears wider than the main jet (Figs~\ref{fig:0206comp}a and b). 
\item The counter-jet has a limb-brightened structure, which is brightest
  between 2.5 and 6\,arcsec from the nucleus, whereas the main jet appears
  narrower and is centrally peaked (Figs~\ref{fig:0206comp}a and b; Figs~\ref{fig:0206trans}a -- f).
\item The longitudinal profile of degree of polarization shows the
  characteristic asymmetry we have noted in other FR\,I jets: the main jet has a
  polarization minimum at $\approx$2.5\,arcsec from the nucleus, corresponding to
  the transition between longitudinal and transverse apparent field, whereas the
  counter-jet shows a high degree of polarization with a transverse apparent
  field, reaching an average of $p \approx 0.5$ at 10\,arcsec (Fig.~\ref{fig:0206polcomp}c). 
\item There is a transition in the field direction between transverse on-axis
  and aligned with the jet boundaries at the edges on both sides of the
  nucleus. This is clear within 2 or 3\,arcsec of the ridge line in the main and
  counter-jets, respectively (Fig.~\ref{fig:0206polcomp} and Figs~\ref{fig:0206trans}j -- o). The signal-to-noise ratio in the data is too low to
  determine the edge field direction accurately at larger distances, so 
  discrepancies between observed and predicted $Q/I$ transverse profiles
  should not be taken too seriously.
\item Close to the nucleus, the apparent field wraps around the edges of both
  jets, with a high degree of polarization, especially on the counter-jet side
  (Figs~\ref{fig:0206polcomp}d and e).
\end{enumerate}

\begin{figure}
\begin{center}
\epsfxsize=7cm
\epsffile{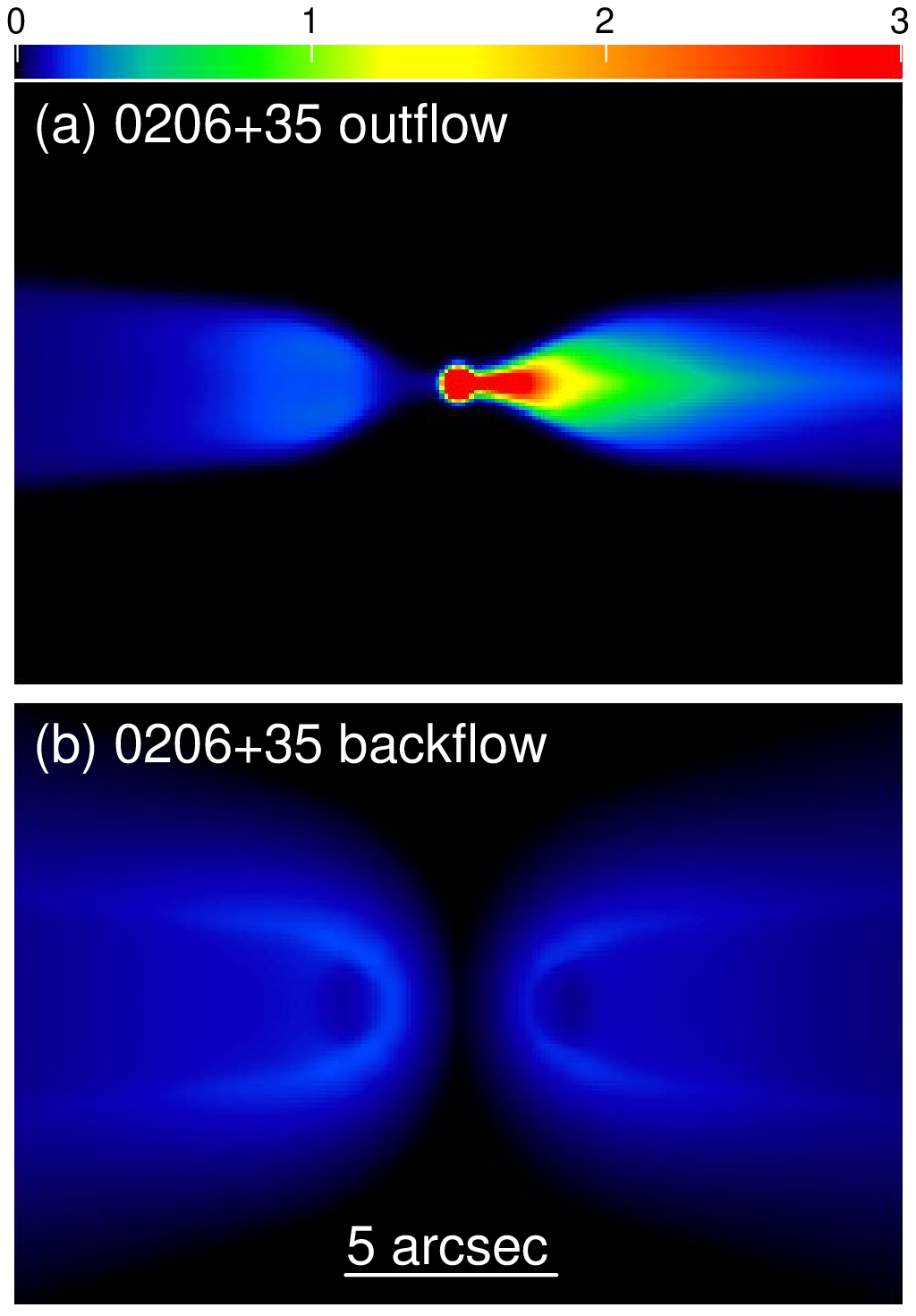}
\caption{Predicted brightness distributions for the outflowing and backflowing
  parts of the model for 0206+35. (a) outflow; (b) backflow.
\label{fig:outback0206}
} 
\end{center}
\end{figure}
The main deficiency of the model is that it underpredicts the brightness of the 
counter-jet $\ga$5\,arcsec from the axis and over-predicts that 
of the main jet between 1.5 and 4\,arcsec. These effects  
lead to a model sidedness ratio which is too high off-axis, although
still significantly $<$1. This discrepancy is most obvious between 5 and 7\,arcsec
from the nucleus (Figs~\ref{fig:0206trans}b, e and h), but is restricted to
regions where the brightness is $\la$200\,$\mu$Jy\,beam$^{-1}$. The model is
also constrained to have monotonic deceleration in the outflow and velocity
independent of distance from the nucleus in the backflow, so it cannot
reproduce the increase in sidedness ratio between 8 and 10\,arcsec from the
nucleus. The surface brightness is low at these distances so uncertainties
in lobe subtraction may be significant.

Fig.~\ref{fig:outback0206} shows the predicted brightness distributions for the
outflowing and backflowing parts of the model separately.  The former is
similar to the pure outflow models we have derived for other
sources 
\citep{LB02a,CL,CLBC,LCBH06}. In the model of 0206+35, the limb-brightening of the counter-jet
is due to a combination of outflow and backflow.  In the outflow, the on-axis 
velocity remains high, so the edges of the outflowing counter-jet material
appear relatively brighter
because they suffer less Doppler dimming than the on-axis material. This effect is 
reinforced by emission from the backflow, which adds a thin shell of emission immediately
surrounding the outflow.  Most of the asymmetry is due to the outflow: the
backflow is only slightly brighter on the counter-jet side.

\subsection{0755+37}
\label{0755fit}

\begin{figure*}
\begin{center}
\epsfxsize=15cm
\epsffile{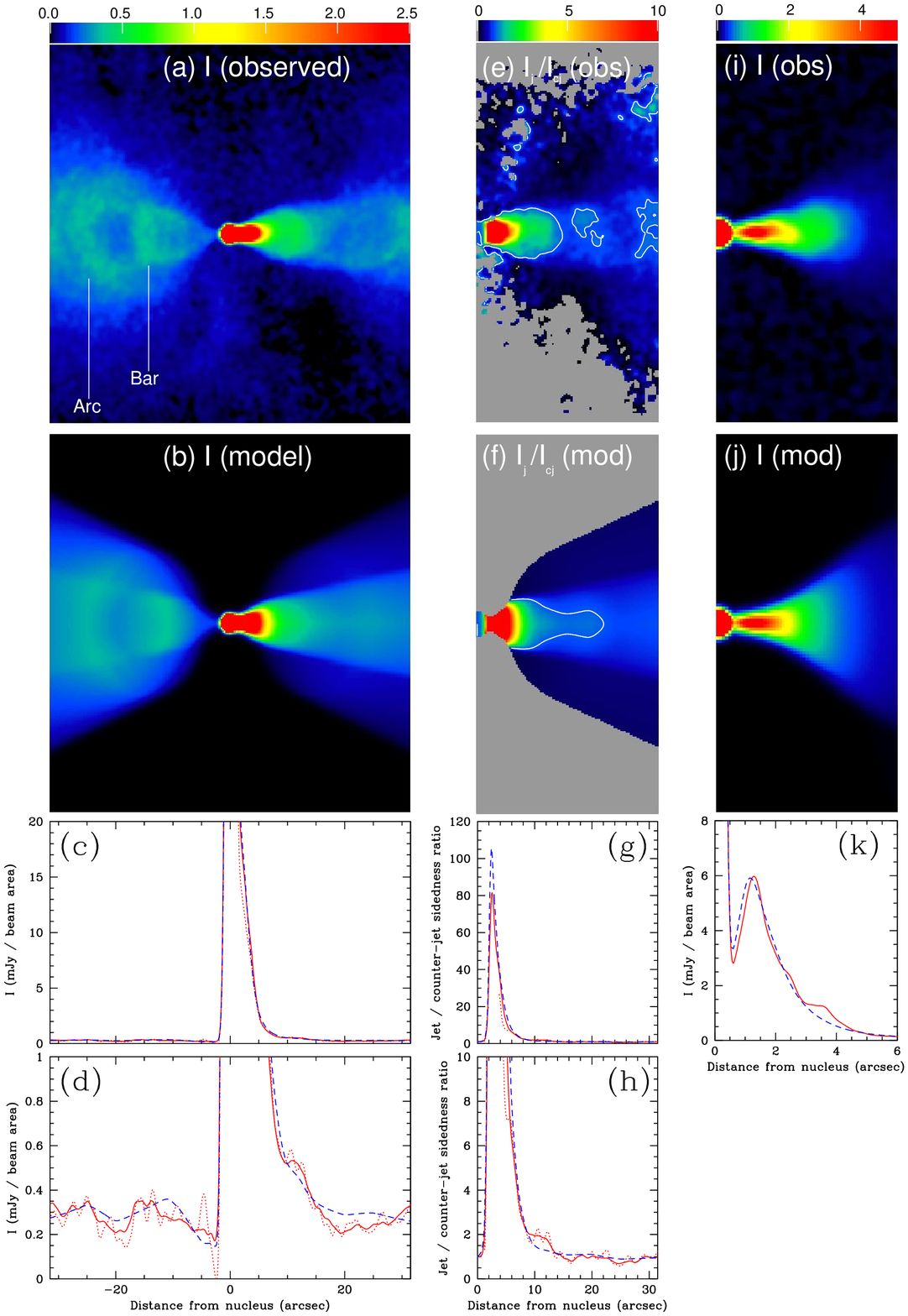}
\caption{Comparison between the observed and modelled total intensities $I$ and
  sidedness ratios $I_{\rm j}/I_{\rm cj}$ for 0755+37. (a) and (b) colour images
  of $I$. (a) observed; (b) model. (c) and (d) profiles of observed and model
  $I$ along the axis of the jet. (e) and (f) images of $I_{\rm j}/I_{\rm
  cj}$.  The white contours represent $I_{\rm j}/I_{\rm cj} =
  1$: outside the contours, $I_{\rm j}/I_{\rm cj} < 1$. (g) and (h) profiles of observed and model $I_{\rm
  j}/I_{\rm cj}$ along the jet axis. The resolution for panels (a) -- (h) is
  1.3\,arcsec FWHM. (i) and (j) colour images of $I$ in the range 0 --
  5\,mJy\,beam$^{-1}$ for the base of the main jet. (i) observed, (j) model. (k)
  profile of observed and model $I$ along the
  jet axis. The resolution for panels (i) -- (k) is 0.4\,arcsec FWHM. In the
  profile plots, the full and dotted (red) lines are from observed images with lobe
  subtraction by interpolation (as for the colour plots) and spectrum,
  respectively. The dashed/blue line is the model.
\label{fig:0755comp}
} 
\end{center}
\end{figure*}

\begin{figure*}
\begin{center}
\epsfxsize=18cm
\epsffile{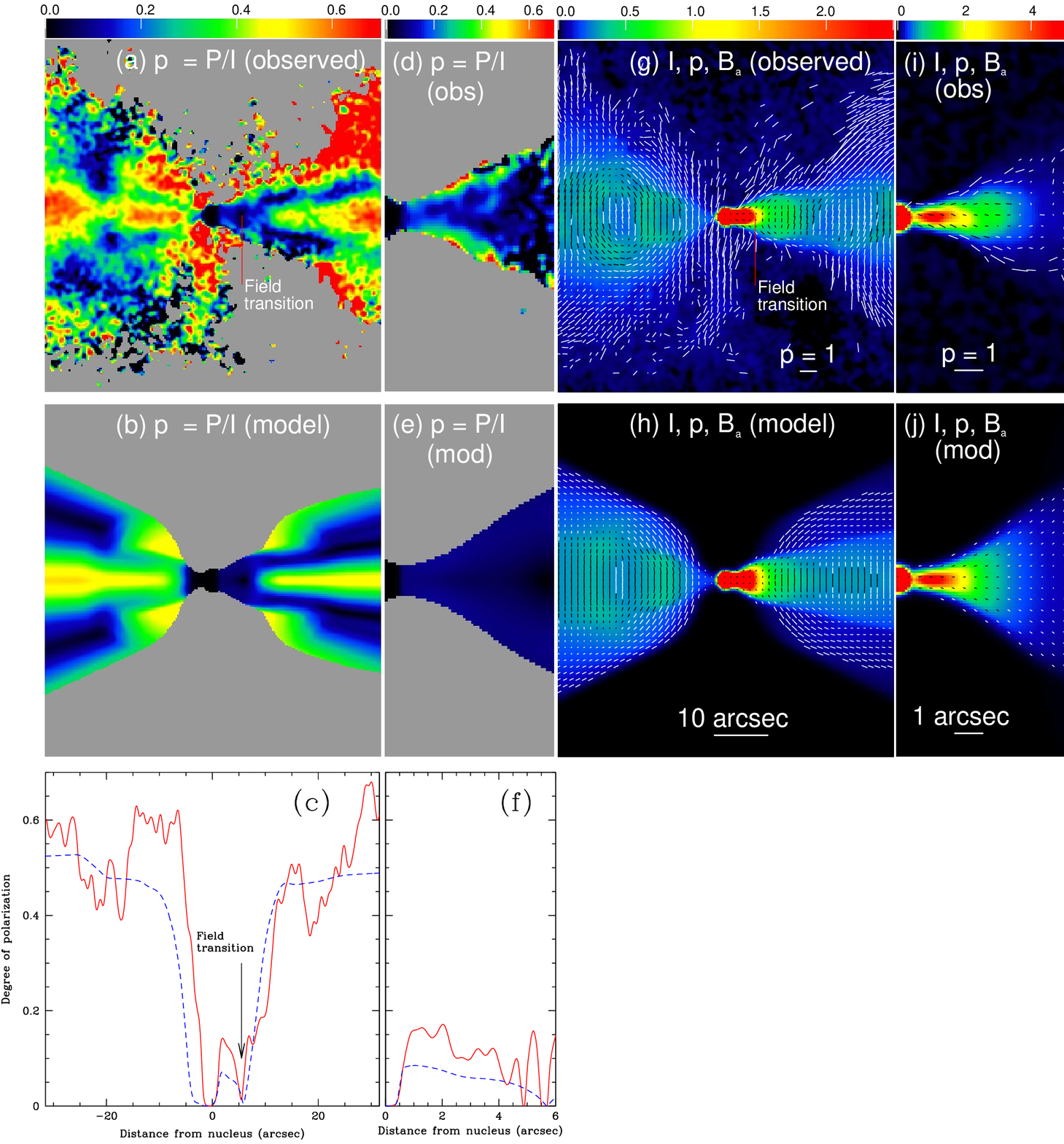}
\caption{Comparison between the observed and modelled linear polarization 
   for 0755+37 at resolutions of 1.3 and 0.4\,arcsec
   FWHM. (a) and (b) colour images
  of $p = P/I$ in the range 0 -- 0.7 at 1.3\,arcsec FWHM. (a) observed; (b) model. (c) 
  profiles of observed (full/red) and model (dashed/blue) $p$ along the axis of the
  jet.  Only the profile of $p$ derived from interpolated images is plotted; the
  equivalent for spectral subtraction is very noisy. (d) -- (f): as (a) -- (c)
  but for the main jet only at 0.4\,arcsec FWHM.
  (g) and (h):  vectors with lengths 
  proportional to $p$ and directions along the apparent magnetic field
  superimposed on colour images of $I$. The resolution is 1.3\,arcsec FWHM and
  the vector scale is indicated by the
  labelled bar. (g) observed, (h) model. (i) and (j): as (g) and (h), but for
  the main jet at 0.4\,arcsec FWHM. 
\label{fig:0755polcomp}
}
\end{center}
\end{figure*}

\begin{figure*}
\begin{center}
\epsfxsize=17cm
\epsffile{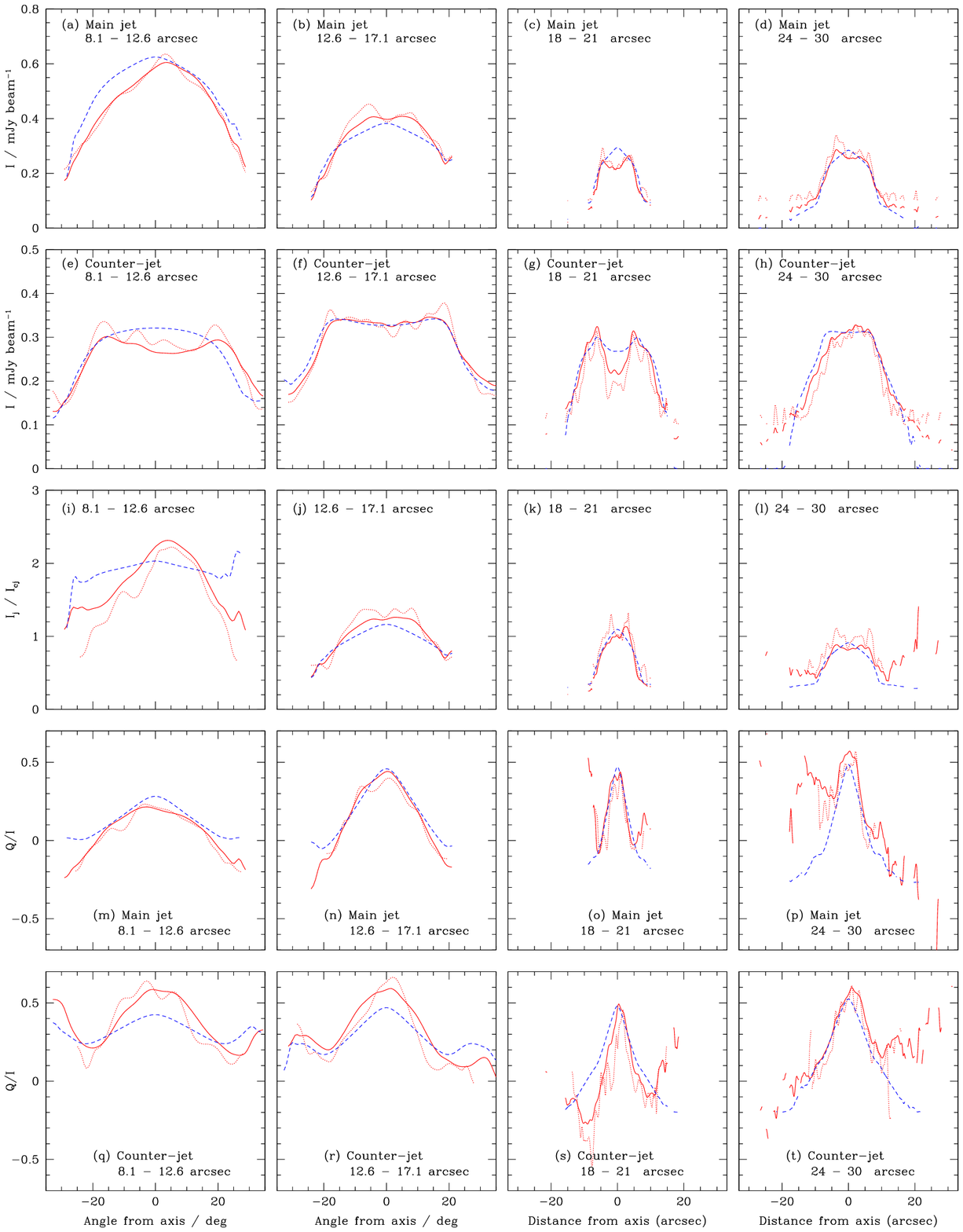}
\caption{Transverse profiles of total intensity, $I$, jet/counter-jet sidedness
  ratio, $I_{\rm j}/I_{\rm cj}$ and $Q /I$ for 0755+37. The data have been
  averaged along radii from the nucleus from 8.1 -- 12.6\,arcsec and from 12.6
  -- 17.1\,arcsec and parallel to the jet axis from 18 -- 21\,arcsec and 24 --
  30\,arcsec, as indicated in the captions (Section~\ref{gencompare}). Full and
  dotted (red) lines both represent observations, with lobe subtraction by
  interpolation and spectral methods, respectively.  Dashed/blue lines show the
  model. $Q/I > 0$ and $Q/I < 0$ correspond to transverse and longitudinal
  apparent field, respectively.
\label{fig:0755trans}
} 
\end{center}
\end{figure*}

We compare model and observed total intensity images and profiles for 0755+37 in
Fig.~\ref{fig:0755comp}; the corresponding polarization comparisons are shown in
Fig.~\ref{fig:0755polcomp} and Fig.~\ref{fig:0755trans} gives averaged
transverse profiles of $I$, $I_{\rm j}/I_{\rm cj}$ and $Q/I$.  Note that the
fainter emission is affected by imperfect lobe subtraction, as discussed in
Section~\ref{images}. This is particularly obvious at large distances from the
jet axis in images of ratios such as $I_{\rm j}/I_{\rm cj}$ and $p$.
The following
features of the brightness and polarization distributions are reproduced.
\begin{enumerate}
\item The main jet has a brightness peak at 1.3\,arcsec from the core
  (Figs~\ref{fig:0755comp}i -- k). Farther out, the profile declines rapidly with
  distance.
\item There is a rapidly-expanding, triangular region of roughly uniform
  brightness at the base of the counter-jet (Figs~\ref{fig:0755comp}a and b).
\item The jet base structure is initially very asymmetric, with a peak sidedness ratio 
  $\approx$80 at 1.9\,arcsec from the core, decreasing rapidly with distance to
  reach an asymptotic value $\approx$1 at 15\,arcsec (Figs~\ref{fig:0755comp}e
  -- h). 
\item  At faint brightness levels, the counter-jet appears significantly
  wider than the main jet, with a large opening angle (Figs~\ref{fig:0755comp}a
  and b).
\item  The counter-jet brightness profiles are more flat-topped or edge-brightened
than those of the main jet at most distances from the core (Figs~\ref{fig:0755comp}a and b and Figs~\ref{fig:0755trans}a -- h).
\item A prominent arc of emission crosses the counter-jet at $\approx$26\,arcsec
  from the nucleus (Figs~\ref{fig:0755comp}a
  and b).
\item There is also a bar of emission crossing the counter-jet at
  $\approx$12\,arcsec from the nucleus (Figs~\ref{fig:0755comp}a, b and d).
\item The profiles of degree and direction of polarization along the axis show
  the same characteristic asymmetry seen in 0206+35 and other FR\,I jets.
  There is a change in apparent field direction at $\approx$5\,arcsec from the
  nucleus in the approaching jet, but not in the counter-jet, whose apparent
  magnetic field is always transverse (Figs~\ref{fig:0755polcomp}g and h, Figs~\ref{fig:0755trans}q -- t). The degree of polarization in the
  counter-jet rises monotonically with distance from the nucleus,
  reaching large values ($p \approx 0.5$) far from the nucleus
  (Figs~\ref{fig:0755polcomp}a -- c).
\item The degree of polarization in the main jet base is low, and the apparent
  field is longitudinal (Figs~\ref{fig:0755polcomp}d -- f, i and j).
\item There are minima in the degree of polarization on either side of the axis
  in both jets, corresponding to the transition between transverse and
  longitudinal apparent field (Figs~\ref{fig:0755polcomp}a, b, g, h;
  \ref{fig:0755trans}m -- t).
\item There is a region of high polarization with a circumferential magnetic
  field around the base of the counter-jet (Figs~\ref{fig:0755polcomp}g and h). 
\item Determination of the observed polarization in the faint
  regions far from the axis is complicated by imperfect subtraction of lobe
  emission, but the apparent field is primarily parallel to the edges of both
  jets (Figs \ref{fig:0755trans}m -- t). 
\end{enumerate}

Features which are not fit well by the model are as follows.
\begin{enumerate}
\item The observed brightness distribution of the bright main jet base is
  slightly more centre-brightened than the model and the observed degree of
  polarization is higher than predicted at its edges (Figs~\ref{fig:0755comp}i,
  j; \ref{fig:0755polcomp}d, e, i, j).
\item The observed transverse total-intensity profiles are significantly more
  limb-brightened than the model in some places, and in particular between 18
  and 21\,arcsec on both sides of the nucleus (Figs~\ref{fig:0755trans}c and g).
\item The inner bar crossing the counter-jet is both straighter and slightly
  farther from the nucleus in the observed image ($\approx$13\,arcsec compared
  with $\approx$11\,arcsec for the model; Figs~\ref{fig:0755comp}a, b, d). The
  fit may be affected by the limb-brightening in this region, however.
\item As in 0206+35, the off-axis brightness of the main jet is slightly overestimated 
  close to the nucleus. The difference is, however, exaggerated by the look-up table in
  Figs~\ref{fig:0755comp}(a) and (b) and is more accurately represented by the
  profile in Fig.~\ref{fig:0755trans}(a). 
\end{enumerate}

\begin{figure}
\begin{center}
\epsfxsize=7cm
\epsffile{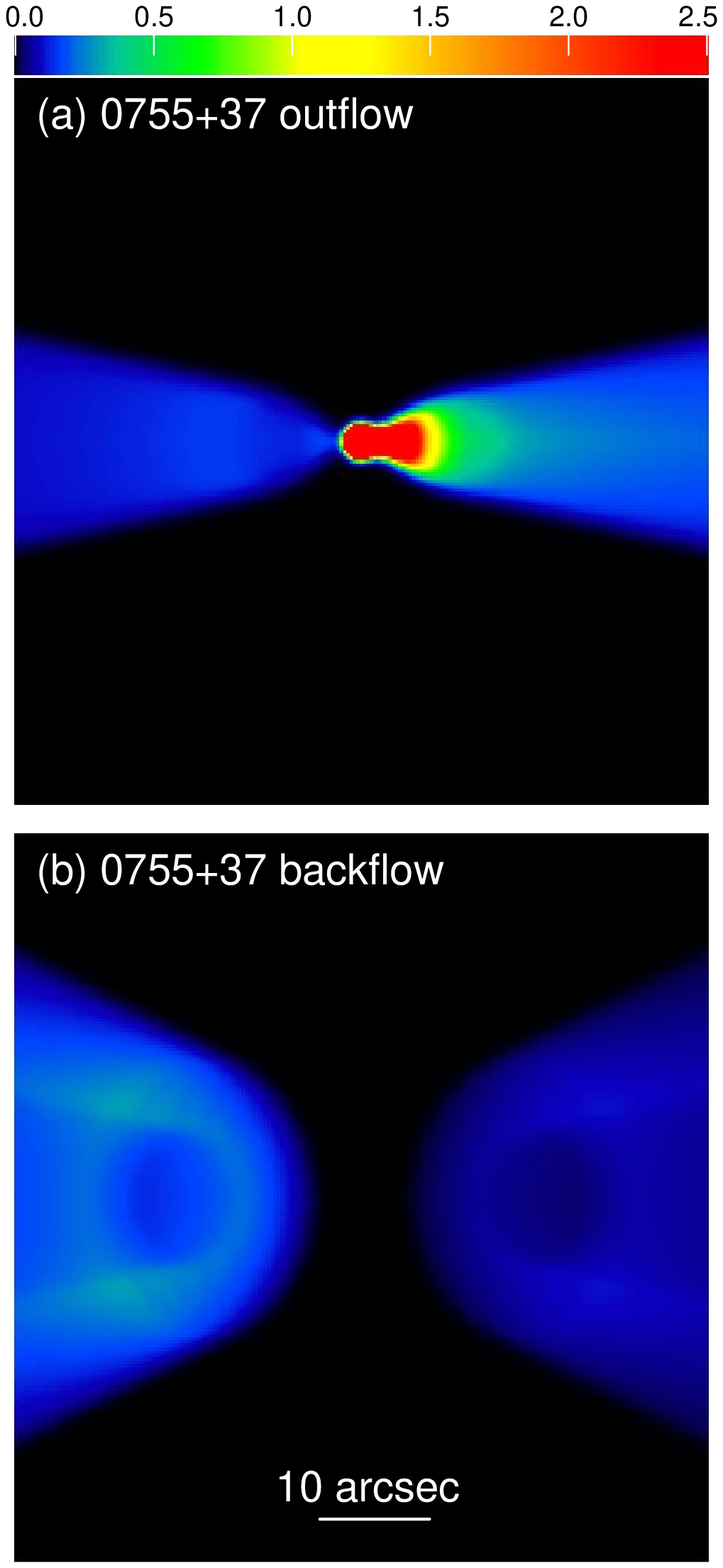}
\caption{Predicted brightness distributions for the outflowing and backflowing
  parts of the model for 0755+37. (a) outflow; (b) backflow.
\label{fig:outback0755}
} 
\end{center}
\end{figure}

Fig.~\ref{fig:outback0755} shows the outflow and backflow components of the
model intensity distribution. As for 0206+35, the outflow appears similar to
that in other FR\,I radio galaxies, but the backflow is relatively stronger in
0755+37. The prominent curved arc crossing the counter-jet $\approx$26\,arcsec
from the nucleus is modelled as the projection of the inner edge of the backflow
at $r = r_{\rm b}$. This is roughly elliptical in shape, with an axial ratio
of $\sec\theta = 1.22$ and there is good correspondence between model and
data. As mentioned earlier, the fit to the bar crossing the counter-jet closer to the nucleus is less
successful. In the model, this is the other half of the projected inner edge of
the backflow, so there is no freedom to adjust its location or curvature to match the
observed feature more closely.   A
similar problem afflicts the main jet: the projection of the inner edge of the
backflow appears slightly too bright, causing the excess off-axis emission close
to the nucleus. 

\section{Derived parameters}
\label{intrinsic}

The best-fitting parameters for our models of 0206+35 and 0755+37 are listed
in Tables~\ref{tab:outflow} (outflow) and \ref{tab:backflow} (backflow).

\subsection{Geometry}

Both sources are fairly close to the line of sight, as expected from their high
jet/counter-jet sidedness ratios and bright cores. We derive $\theta = 40^\circ$
for 0206+35 and $35^\circ$ for 0755+37.  The outflow geometries are typical of
those we have determined for other FR\,I jets, with the boundaries between
flaring and outer regions at 5.3 and 13.9\,kpc from the nucleus for 0206+35 and
0755+37, respectively. The corresponding half-opening angles in the outer
regions are 3\fdg 9 and 7\fdg 4. 

In 0206+35, the backflow has a half-opening angle of 11\degr\ in the outer
region and its emission extends back into the flaring region, with a cut-off at $r_{\rm b}
= 2.7$\,kpc. For 0755+37, on the other hand, the backflow emission is truncated within
the outer region ($r_{\rm b} = 23$\,kpc), where its half-opening angle is
16\degr.

\subsection{Velocity}
\label{vel-parms}

Velocity images derived from our model fits are shown in Fig.~\ref{fig:vels}.

The initial velocities of both outflow components are similar ($\beta_1 = 0.86$
for 0206+35 and 0.88 for 0755+37) and the associated transverse velocity
profiles are close to uniform. 0206+35 shows little on-axis deceleration,
reaching an asymptotic velocity $\beta_0 = 0.68$ after 4\,kpc. Its transverse
velocity profile evolves much more, and the fractional edge velocity is 0.04 at
large distances.  In both these respects, the source resembles 3C\,296
\citep{LCBH06}. 0755+37, on the other hand, appears to decelerate rapidly, to
$\beta_0 = 0.25$ by 18.5\,kpc, with a fractional edge velocity of 0.26. This
estimate should be treated with caution since the emission in the outer
counter-jet is dominated by the backflow component, making it difficult to
assess the intensity or polarization of the outflow there.

The backflow velocities increase away from the source axis, from $\beta = 0.05$
to 0.20 for 0206+35 and from 0.25 to 0.35 for 0755+37. 

\subsection{Emissivity}

Model images of $n_0 B^{1+\alpha}$ (proportional to the emissivity function
$\epsilon$) are shown in Fig.~\ref{fig:emiss}.

The model outflow components again show properties very similar to those in
other FR\,I jets.  The locations of the flaring points (0.82 and 1.55\,kpc from
the nucleus for 0206+35 and 0755+37, respectively) are well determined and
consistent with higher-resolution observations \citep{LGBPB}.  The emissivity
variations in the faint and poorly resolved inner jets upstream of the flaring
points are not well constrained. In the flaring and outer regions, the gradient
of the emissivity profile flattens with distance in both sources, as is usual in
FR\,I jets. 0755+37 requires a sudden decrease in emissivity with distance at $r
= r_{e0}$ whereas 0206+35 does not. 

The observed limb-brightening in both sources shows side-to-side symmetry.  This
cannot result from a transverse velocity gradient in the sense we have inferred,
which would lead to limb-brightening only in the counter-jet.  In agreement with
this qualitative argument, the best-fitting transverse emissivity profiles are
higher at the edges than on-axis. This effect is slight in 0206+35, where the
profile is consistent with a uniformly-filled cylinder everywhere. In 0755+37,
however, limb-brightening is required over much of the outer region
(Fig.~\ref{fig:emiss}b).  As noted in Section~\ref{0755fit}, the observed
transverse intensity profiles in this source are significantly more
limb-brightened than the model predicts, suggesting that there is a narrow
enhancement in emissivity at the boundary between the outflow and backflow. The
functional form we assume for the transverse variation of emissivity does not
allow for such narrow features.

The backflow emissivity decreases with distance at similar rates in the two
sources ($\propto r^{-1.66}$ in 0206+35 and $\propto r^{-1.81}$ in 0755+37).  It
is centre-brightened in 0206+35 ($e_{\rm b} = 0.02$) but closer to uniform in
0755+37 ($e_{\rm b} = 0.79$).

\subsection{Field Ordering}

The fractional components of magnetic field, $\langle B^2_t \rangle^{1/2}/B$
(toroidal), $\langle B^2_l \rangle^{1/2}/B$ (longitudinal), and $\langle B^2_r
\rangle^{1/2}/B$ (radial) are plotted in Fig.~\ref{fig:bfield}.

In both sources, the field close to the nucleus in the outflow is close
to isotropic, with the longitudinal component just exceeding the other
two. At larger distances, the toroidal component dominates, with significant
longitudinal and radial contributions in 0206+35 and 0755+37, respectively. As
for velocity and emissivity, the field components in the outer parts of 0755+37
may have larger systematic errors because of the dominance of backflow emission.

The field in the backflow is toroidally dominated in both sources, with
non-negligible radial components in both cases and some longitudinal field in
0206+35.  
  
\begin{center}
\begin{table}

\caption{Model parameters which are common to outflow and backflow, or which
  apply only to the outflow (Section~\ref{fit-funcs} and Table~\ref{tab:functions}). Col.\,1: parameter; col.\,2: unit;
  cols 3 and 4: values for 0206+35 and 0755+37. The parameters are defined in
  Section~\ref{fit-funcs} and listed in
  Table~\ref{tab:functions}. $\Delta\theta$ is the range of angles to the line
  of sight for which any acceptable solutions can be obtained.\label{tab:outflow}}
\begin{tabular}{lrrr}
\hline 
&&&\\
\multicolumn{2}{c}{Variable} &0206+35&0755+37\\ 
&&&\\
\hline
&&&\\
\multicolumn{4}{c}{Geometry (common to outflow and backflow)}\\
&&&\\
$\theta$&deg      &$40.0_{-0.3}^{+0.3}$  &$34.8_{-0.8}^{+0.7}$  \\
$\Delta\theta$&deg&$34 - 43$             &$32.5 - 37.5$         \\
$r_0$&kpc         &$5.3_{-0.1}^{+0.1}$   &$13.9_{-0.3}^{+0.3}$  \\
&&&\\
\multicolumn{4}{c}{Outflow geometry}\\
&&&\\
$\xi_0$&deg       &$3.9_{-0.2}^{+0.2}$   &$7.4_{-0.1}^{+0.2}$   \\
$x_0$&kpc         &$1.32_{-0.04}^{+0.02}$&$3.88_{-0.06}^{+0.08}$\\
&&&\\
\multicolumn{4}{c}{Velocity}\\
&&&\\
$r_{v1}$&kpc&$1.8_{-0.3}^{+0.3}$   &$3.6_{-1.5}^{+1.6}$   \\
$r_{v0}$&kpc&$4.1_{-0.2}^{+0.3}$   &$18.5_{-1.5}^{+2.3}$  \\
$\beta_1$&  &$0.86_{-0.07}^{+0.08}$&$0.88_{-0.04}^{+0.05}$\\
$\beta_0$&  &$0.68_{-0.05}^{+0.09}$&$0.25_{-0.05}^{+0.07}$\\
$v_1$&      &$0.95_{-0.13}^{+0.05}$&$1.00_{-0.06}        $\\
$v_0$&      &$0.04_{-0.01}^{+0.02}$&$0.26_{-0.11}^{+0.19}$\\
&&&\\
\multicolumn{4}{c}{Emissivity}\\
&&&\\
$r_{e1}$&kpc  &$0.82_{-0.02}^{+0.02}$&$1.55_{-0.03}^{+0.04}$\\
$r_{e0}$&kpc  &$2.04_{-0.06}^{+0.07}$&$10.2_{-0.3}^{+0.1}$  \\
$E_{\rm in}$& &$\approx 3.1$         &$\approx 2.4$         \\
$E_{\rm mid}$&&$2.59_{-0.08}^{+0.09}$&$3.76_{-0.04}^{+0.02}$\\
$E_{\rm out}$&&$2.13_{-0.06}^{+0.08}$&$1.16_{-0.09}^{+0.05}$\\
$e_1$&        &$1.2_{-0.5}^{+0.6}$   &$1.0_{-0.2}^{+0.3}$   \\
$e_0$&        &$1.14_{-0.16}^{+0.16}$&$2.2_{-0.3}^{+0.5}$   \\
$g_1$&        &$1.7_{-1.3}^{+0.8}$   &$1.7_{-0.4}^{+0.5}$   \\
$g_0$&        &$1.05_{-0.09}^{+0.08}$&$0.52_{-0.03}^{+0.06}$\\
&&&\\
\multicolumn{4}{c}{Field component ratios}\\
&&&\\
$r_{B1}$&kpc     &$<1.4$                &$8.8_{-2.0}^{+2.8}$   \\
$r_{B0}$&kpc     &$4.6_{-0.5}^{+0.5}$   &$15.4_{-3.2}^{+2.5}$  \\
$j_1$ &&$1.50_{-0.22}^{+0.34}$&$0.96_{-0.09}^{+0.13}$\\
$j_0$ &&$0.11_{-0.11}^{+0.13}$&$0.44_{-0.15}^{+0.12}$\\
$k_1$ &&$1.36_{-0.13}^{+0.13}$&$1.15_{-0.07}^{+0.08}$\\
$k_0$ &&$0.64_{-0.04}^{+0.05}$&$0.08_{-0.08}^{+0.22}$\\
&&&\\
\hline
\end{tabular}
\end{table}
\end{center}

\begin{center}
\begin{table}
  \caption{Model parameters for backflow (Section~\ref{fit-funcs} and Table~\ref{tab:functions}).  Col.\,1: parameter; col.\,2: unit;
    cols 3 and 4: values for 0206+35 and 0755+37.\label{tab:backflow}}
\begin{tabular}{lrrr}
\hline 
&&&\\
\multicolumn{2}{c}{Variable} &0206+35&0755+37\\ 
&&&\\
\hline
&&&\\
\multicolumn{4}{c}{Geometry}\\
&&&\\
$\xi_{\rm b}$&deg&$10.9_{-0.5}^{+0.5}$&$15.6_{-0.1}^{+0.5}$\\
$r_{\rm b}$&kpc  &$2.7_{-0.2}^{+0.1}$ &$23.2_{-0.7}^{+0.8}$\\
&&&\\
\multicolumn{4}{c}{Velocity}\\
&&&\\
$\beta_{\rm b, in}$ &&$0.02_{-0.02}^{+0.03}$&$0.25_{-0.07}^{+0.04}$\\
$\beta_{\rm b, out}$&&$0.20_{-0.07}^{+0.06}$&$0.35_{-0.05}^{+0.05}$\\
&&&\\
\multicolumn{4}{c}{Emissivity}\\
&&&\\
$n_{\rm b}$&$\times 100$&$2.3_{-0.2}^{+0.2}$   &$0.094_{-0.010}^{+0.000}$\\
$E_{\rm b}$&            &$1.66_{-0.07}^{+0.06}$&$1.81_{-0.05}^{+0.07}$\\
$e_{\rm b}$&            &$0.05_{-0.01}^{+0.02}$&$0.79_{-0.14}^{+0.13}$\\
&&&\\
\multicolumn{4}{c}{Field component ratios}\\
&&&\\
$j_{\rm b}$&&$0.24_{-0.07}^{+0.08}$&$0.38_{-0.07}^{+0.07}$\\
$k_{\rm b}$&&$0.38_{-0.09}^{+0.08}$&$0.03_{-0.03}^{+0.15}$\\
&&&\\
\hline
\end{tabular}
\end{table}
\end{center}

\begin{figure}
\begin{center}
\epsfxsize=6.0cm
\epsffile{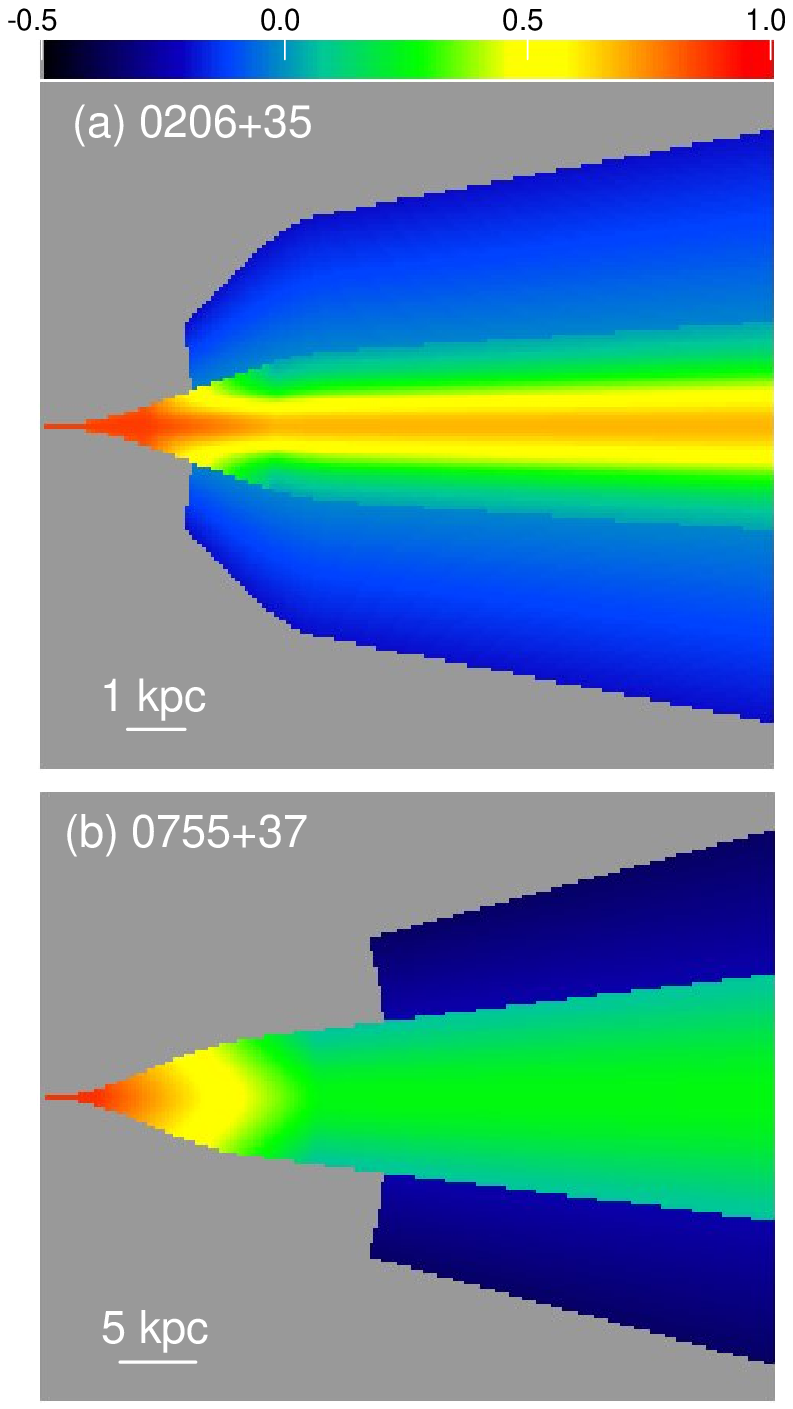}
\caption{The model values of velocity $\beta$ in units of $c$ in planes
  containing the jet axes. Positive and negative values of $\beta$ denote
  outflow and backflow, respectively. (a) 0206+35, (b) 0755+37.
\label{fig:vels}
}
\end{center}
\end{figure}

\begin{figure}
\begin{center}
\epsfxsize=6.0cm
\epsffile{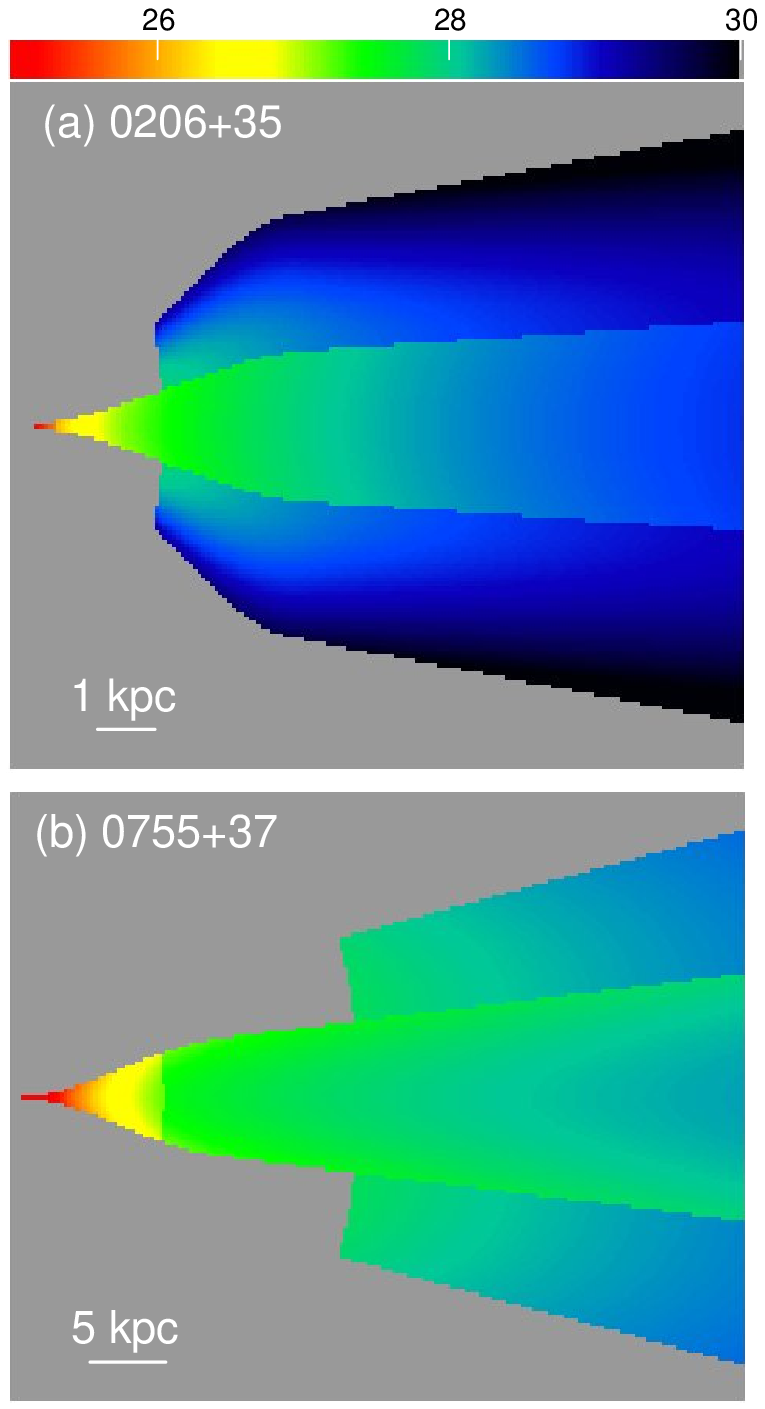}
\caption{The model values of $\log (n_0 B^{1+\alpha})$ in planes containing the
  jet axes ($n_0$ and $B$ are in SI units). (a) 0206+35, (b) 0755+37.
\label{fig:emiss}
}
\end{center}
\end{figure}

\subsection{Backflow spectral index}
\label{spectra}

We can also constrain the spectral index of the radio emission from
the backflows.  The spectral indices at the edges of the
jets, where the line of sight is mainly through backflow emission after the lobe
subtraction, are much closer to those of the jets themselves than to the values
elsewhere in the lobes. We can estimate the spectrum of the backflow emission
directly from the images shown in Figures~\ref{fig:0206sub}(c) and
\ref{fig:0755sub}(c) or, more accurately, by integrating total intensity at
1.425 and 4.860\,GHz over pixels which are unblanked in these images. The latter
method gives mean spectral indices of 0.50 for 0206+35 and 0.57 for 0755+37,
compared with 0.55 and 0.53 for the sum of outflow and backflow emission.

\section{Discussion}
\label{discuss}

\subsection{Testing the hypothesis}

It is clear that the initial jet base asymmetries of most FR\,I jets are
produced by relativistic aberration \citep{LB02a,CL,CLBC,LCBH06}.  If 0206+35
and 0755+37 prove to be typical -- in that counter-jets consistently appear
wider than the main jets at a given isophote in lobed FR\,I sources whose jet
base asymmetries are large -- then the jet/counter-jet width asymmetry must also
be correlated with jet orientation.  The models presented in
Section~\ref{compare} show that mildly relativistic backflow offers a possible
cause for such an orientation-dependent effect.  

There is an alternative explanation 
for $I_{\rm j}/I_{\rm cj}$ becoming $<$1 in some parts of a source
which 
also
preserves the orientation-dependence of the effect.
For the special case where the
magnetic field is
purely toroidal and the edge velocity is $\approx \cos\theta$, it is possible
for relativistic aberration to give an off-axis jet/counter-jet sidedness ratio
$<$1 even for a pure symmetrical 
{\em outflow}. 
We analyse this 
special
case in
Appendix~\ref{toroidal}, where we show that it is {\em inconsistent} with the
polarization imaging of 0206+35 and 0755+37.  The mechanism inevitably produces
degrees of polarization close to the theoretical maximum of $p_0 \approx 0.7$
with a transverse apparent field.
It is therefore unlikely to be important in
the majority of observed jets 
but it 
may be relevant in a few objects like 3C\,296
(Appendix~\ref{toroidal}).

If the jets are intrinsically symmetrical, then the backflow hypothesis remains
the most plausible explanation for the observed brightness and polarization
asymmetries, but (with only two clear-cut cases analysed in such detail so far)
it is important to test it by looking at more objects.  We reviewed the rest of
the B2 low-luminosity source sample \citep{Parma87} to see if any other data
support (or contest) the interpretation given here.  \cite{LPdRF} found that the
source B2\,0844+31 also has both a small jet to counter-jet width ratio and a
high intensity ratio $I_{\rm j}/I_{\rm cj}$.  Unfortunately, there is no imaging
for that source of the high quality we now have for 0206+35 and 0755+37 so we
cannot test models of its asymmetries at the same level of detail.  Nor can we
classify its large scale structure definitively as `lobed' or `plumed': deeper
imaging sensitive to its most extended structure is needed.  Although lack of
high-quality imaging precludes us from finding other good examples of these
phenomena in the B2 sample, we note that there are no clear {\sl
  counter}-examples -- either of sources in which the brighter jet appears to be
wider than the counter-jet at low intensity levels, or of a large
jet/counter-jet width asymmetry in a source that lacks `lobed' structure or with
only a small jet/counter-jet {\sl intensity} asymmetry at its base.

As noted by \citet[see their fig.~15]{LCBH06}, the jets in the lobed FR\,I
source 3C\,296 show $I_{\rm j}/I_{\rm cj} < 1$ at their edges. The emission
there is faint, but the effect is consistently present in the flaring and outer
regions. The transverse variations of linear polarization are also very
different in the two jets \citep[Figs~18g and h]{LCBH06}: the counter-jet shows
a prominent parallel-field edge, whereas the main jet does not.  The model
described by \citet{LCBH06}, while giving a good overall fit to the brightness
and polarization distributions of 3C\,296, was not consistent with the
observation of $I_{\rm j}/I_{\rm cj} < 1$ and did not fully reproduce the flat
profile of $p$ with transverse apparent field in the approaching jet.  We have
examined possible backflow models for 3C\,296 and find that they are
qualitatively inconsistent with the polarization distribution, although they can
easily fit the edge sidedness ratios. The combination of sidedness ratio and
polarization is more reminiscent of the predictions of the outflow model
analysed in Appendix~\ref{toroidal}.

Emission from backflow such as that modelled here would be hard to recognise 
in lobed FR\,I sources whose jets are close to the plane of the sky.  The backflow emission in
such sources would be almost indistinguishable from faint outer edges of their jets and
only unusually precise spectral index measurements could distinguish it from low level 
brightness enhancements of the lobes near the jets.

\begin{figure*}
\begin{center}
\epsfxsize=15cm
\epsffile{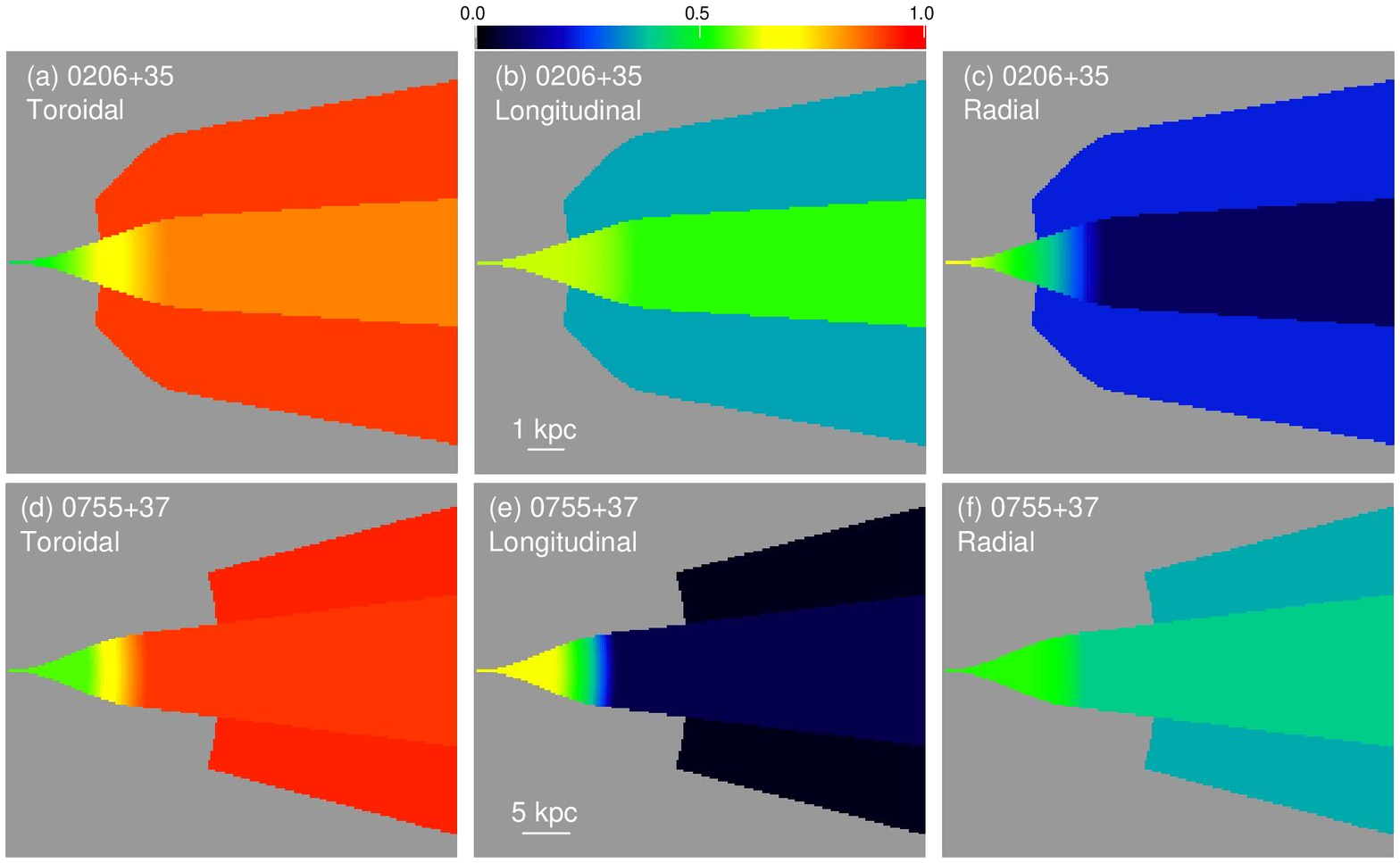}
\caption{The fractional magnetic field components for the three sources. (a),
  (d) toroidal, $\langle B_t^2/B^2\rangle^{1/2}$; (b), (e) 
  longitudinal, $\langle B_l^2/B^2\rangle^{1/2}$; (c), (f) radial, $\langle
  B_r^2/B^2\rangle^{1/2}$.  (a) -- (c) 0206+35, (d) -- (f) 0755+37.
\label{fig:bfield}
} 
\end{center}
\end{figure*}

\subsection{Should we expect backflows in FR\,I sources?}

Light jets propagating into dense media can be expected to terminate in one of
two ways. They may decelerate and transition into `plumes' or `tails' that are
deflected away from the AGN by external pressure gradients or by winds in the
IGM.  Alternatively, they may deflect before reaching a contact discontinuity
with the denser external medium, thus accumulating a `cocoon' around the
outflow.  The first process is thought to underlie the formation of plumed or
tailed FR\,I radio sources such as 3C\,31 while the second is thought to form
the `classical double' lobed radio sources such as Cygnus A and is often
associated with FR\,II morphology.  Lobes in the (generally more luminous)
FR\,II sources also frequently contain discrete radio `hot spots' that are
identified with strong shocks where well-collimated (supersonic) outflows are
slowed and begin to supply lobe material. There is no reason to suppose,
however, that discrete hot spot formation is a requisite for cocoon (or radio
lobe) formation -- momentum balance alone requires the deflection of the light
outflow if it cannot escape along its initial path owing to development of a
high pressure region downstream.  Cocoons without hot spots are indeed seen in
simulations of relativistic jets which are much lighter than their surroundings
\citep{PM07,Rossi08}, in which the jets flows are transonic where they terminate.

The majority of FR\,I sources form radio lobes 
whose detailed morphologies, spectral characteristics and polarization properties strongly 
resemble those of higher-power FR\,II lobes \citep{PDF96,Parma99,LGBPB}. Their lobes
have sharp outer brightness gradients, 
circumferential magnetic fields, and spectral indices that steepen towards the centre 
of the source on the largest angular scales -- but {\sl without} hot spots.  Furthermore,
 outflows in lobed FR\,I sources can deflect 
through large angles without losing their identities: \cite{LGBPB} found  regions
where emission with jet-like spectral index $\approx$0.6 had displaced steeper-spectrum 
emission within FR\,I lobes. These results suggest that ongoing large-scale flow is 
present in these lobes well beyond the clearly recognisable jets.
  
There is therefore both theoretical and observational support for
supposing that jet outflows containing relativistic particles and
magnetic fields may be redirected through large angles in lobed FR\,I
sources.  The additional ingredient suggested by our modelling of
0206+35 and 0755+37 is that a component of such an outflow in an FR\,I
source can return to the vicinity of the AGN as mildly relativistic
backflow.  As we noted in the introduction to this paper, this idea is
supported by the presence of backflow with $\beta \ga 0.2$ around the
jets in some numerical simulations of the propagation of light,
relativistic jets.  The simulation by \citet{PM07} used initial
conditions for the jet derived from our FR\,I source models
\citep{LB02a,LB02b} and realistic density and pressure gradients in
the surrounding galactic and group atmosphere \citep{Hard02}. In
particular, the velocity at injection was $\beta = 0.87$ and the
initial density contrast (the ratio of the density of the jet to that
of its surroundings) was $\eta = 10^{-5}$.  Although the jet had
propagated only $\approx$15\,kpc by the end of the simulation, the
structure already resembled a lobed FR\,I source of the type discussed
here, with a cocoon of backflowing, mixed jet and external plasma
surrounding the jet. The jet was transonic at its termination, so no
hot spot was formed.  Typical backflow velocities in the cocoon were
$\beta \approx 0.15$, with values reaching $\beta \approx 0.4$ close
to the nucleus.  The use of an open boundary condition in the symmetry
plane at the base of the jet can cause the backflow speed to be
over-estimated \citep{Saxton02}, although \citet{PM07} argued that this
effect was small in their simulation because the flow through the open
boundary was negligible. One other possible concern is that the
simulation by \cite{PM07} was axisymmetric: the speed and extent of
fast backflow appear to be smaller in some fully three-dimensional
simulations compared with the equivalent axisymmetric cases
\citep{Norman96,Aloy99}. We note, however, that the comparison may not
be relevant to lobed FR\,I sources because the density contrast, $\eta
= 0.01$, was much higher in these two examples, leading to cocoons
which were far longer and thinner than those observed. The
three-dimensional simulation of a relativistic jet with $\eta =
10^{-4}$ by \citet{Rossi08} indeed showed fast backflow with $\beta
\approx 0.4$, despite the use of symmetric boundary conditions at the
jet inlet. The initial conditions (jet Lorentz factor $\Gamma = 10$)
and the assumption of a uniform external density are probably more
appropriate to smaller physical scales than we consider here, however.
Thus, although the assumptions and initial conditions of the
simulations by \citet{PM07} and \cite{Rossi08} are not realistic
enough to permit a quantitative comparison with our results, they do
suggest that the idea of fast backflow is a reasonable one provided
that the density contrast is very small ($\la 10^{-4}$).

The simulations discussed above are entirely hydrodynamic.  We also note that
backflow is an expected ingredient of models of magnetic hoop stress collimation
of current-carrying jets because such models must provide a return current path --
although it is unclear that such return paths need be as close to the jet outflow
boundary as the backflow we have described here.

\section{Summary and Further Work}
\label{summary-further}

\subsection{Summary}
\label{summary}

We have shown that many aspects of the intensity and linear polarization distributions
over the inner jets and counter-jets in the lobed FR\,I radio sources 0206+35 and
0755+37 are accounted for by an intrinsically symmetrical decelerating 
relativistic jet model that includes (mildly) relativistic backflow around both jets.

We have estimated properties of this backflow subject to the simplifying assumptions 
that it is symmetrical across the AGN, axisymmetric, and that its streamlines are 
similar in shape to those of the outflow.   Although these assumptions are likely
to be too simple a priori we nevertheless find that the quality of the $IQU$ fits 
obtained with the models including such symmetric backflow is 
similar to that obtained with pure decelerating outflow models of other 
FR\,I jets \citep{LB02a,CL,CLBC,LCBH06}.   Furthermore, the outflow components of
the models we have fitted to 0206+35 and 0755+37 are quite similar to those 
obtained for other FR\,I sources.  The addition of backflow to the models 
therefore suffices to explain the otherwise anomalous jet/counter-jet
asymmetries of both sources and eliminates the need to invoke ad hoc
environmental (or other intrinsic) side-to-side asymmetries.   

The salient features of backflow inferred from this procedure are as follows.
\begin{enumerate}
\item The backflow velocities are mildly relativistic, in the range $0.05 \la \beta \la 0.35$ (Fig.~\ref{fig:vels}).
\item The backflows are approximately symmetric around the outflows and their radio 
emission comes from a hollow cone surrounding the jet axis with additional half-opening
angles $\approx 8\degr$. 
\item  They can be traced to considerable distances from the AGN (at least 15\,kpc
for 0206+35 and 50\,kpc for 0755+37) but the emission close to the ends of the
jets in both sources is chaotic, and it is not clear where the backflows
begin.
\item They do not emit synchrotron radiation all the way in to the AGN (Fig.~\ref{fig:emiss}). 
\item The backflows emit with a spectral index $\alpha \approx 0.55$
  (Section~\ref{spectra}). This spectral index is lower than that of the
  nearby lobes and comparable with those of the outflows.
\item  Their magnetic fields are mostly toroidal and their emissivities decrease
with distance roughly as $r^{-1.7}$ (Figs~\ref{fig:emiss} and \ref{fig:bfield}).
\end{enumerate}
 
These are the only two lobed FR\,I sources for which we have deep enough imaging and 
polarimetry to reveal the `two-component' aspect of the jets and counter-jets that 
motivated this study. The generality of our results could thus be called into question by a 
{\sl single} new example of an FR\,I source with strong jet-width asymmetries in which 
either (a) the axis is inferred to be close to the plane of the sky or (b) the apparently wider features are 
associated with the brighter jet.  With only two examples of possible backflow features
we also cannot address whether {\sl all} lobed FR\,I sources might contain backflow or 
(conversely) whether backflow exists {\sl only} in lobed sources.

The interpretation including backflow will continue to
be preferable to any involving intrinsic side-to-side width differences if further studies 
find the apparently wider features only on the counter-jet side, and only in lobed sources 
for which inclination indicators suggest that the jets are at moderately large angles to the 
plane of the sky. 

\subsection{Open questions and further work}
\label{open}

Our observations and models give no clue about the ultimate fate of
the backflow or how it may interact with the outflow, but they raise a
number of questions which could be addressed by deeper,
higher-resolution observations of 0206+35 and 0755+37.
\begin{enumerate}
\item  Where does the backflow originate? Does it start in a high-pressure
  region at the end of the outflow? 
\item  Does the backflow shield the jet from entrainment or interaction with the
  lobe plasma? 
\item  Does the presence of the backflow perturb the jet structure in any way? 
\item  Where does the backflow ultimately go: sideways or even closer to the AGN?
\item  Why does the backflow radiate strongly where it does and stop radiating
  close to the AGN?
\item Can the backflow really be faster than the asymptotic velocity of the
  outflow, as appears at first sight to be the case in 0755+37\footnote{The
    asymptotic outflow velocity is poorly constrained (Section~\ref{vel-parms}),
    so this difference may not be real.}?
\end{enumerate}

Additional questions which could be answered by observations of a
sample of FR\,I sources include the following.
\begin{enumerate} 
\item Do jets in other FR\,I sources with large jet-width asymmetries
  also have the two-component jet and counter-jet structure found here
  in 0206+35 and 0755+37 (i.e.\ a strongly centrally brightened peaked
  main jet and centre-darkened counter-jet near the axis, and
  counter-jet emission consistently brighter than that of the main jet
  further from the axis)?
\item Does the counter-jet/jet width asymmetry indeed correlate well
  with orientation indicators -- counter-jet/jet intensity ratios and
  normalized core power -- as expected in a relativistic backflow
  model of this asymmetry?  If so, the tightness of the correlation
  with orientation indicators could be used to constrain the intrinsic
  symmetry of the backflow.
\item Does the width asymmetry indeed occur only in {\sl lobed}
  FR\,I's?  It will be important to obtain images which are
  sufficiently sensitive to extended structure to detect faint lobe
  emission in any sources whose structural classification is dubious.
\end{enumerate}
The high sensitivity and resolution of the imaging needed to address
all of these issues and to test backflow models of the type we have
proposed will require the use of the Jansky (Expanded) Very Large
Array and {\sl e-MERLIN}.

Given the similarities between the extended emission in FR\,II and
lobed FR\,I sources, it would also be interesting to search for
evidence of backflow in the former class. The jets in FR\,II sources
are usually much narrower than those we have imaged in the present
study and are thought to be highly supersonic where they terminate in
compact hot spots.  Backflow is predicted by simulations of FR\,II
dynamics, but it is unclear how its properties might depend on density
contrast, Mach number, magnetization and source age.  It may be that
observations of FR\,II sources without prominent hot spots will offer
the best chance of detecting backflows. Counter-jets in FR\,II sources
are faint and difficult to distinguish from filamentary lobe emission,
so identification of any backflow component may be even more
challenging than in FR\,I's.

Three-dimensional simulations of very light, relativistic jets
propagating in realistic external density and pressure distributions
would be extremely valuable in understanding the backflow
phenomenon in FR\,I sources. To be realistic, such simulations should
be bipolar, with initial density contrasts $\approx$10$^{-5}$. The
effects of magnetic fields (ordered or disordered) on the flow also
remain to be investigated.

\section*{Acknowledgements}

The National Radio Astronomy Observatory is a facility of the National Science
Foundation operated by Associated Universities, Inc. under co-operative
agreement with the National Science Foundation.  We are grateful to the referee
for a very careful reading of the paper. RAL would also like to thank Alan and
Mary Bridle for hospitality.

\appendix

\section{Pure outflow models with $I_{\rm j}/I_{\rm cj} < 1$}
\label{toroidal}

It is possible under some circumstances for the ratio $I_{\rm j}/I_{\rm cj}$
(approaching/receding) to be significantly less than unity close to the edges of
the brightness distribution even for a symmetrical {\em outflow}.  This might
easily be mistaken for the effects of a backflowing component. We argue in this
Appendix that the effect is quite likely to be observed in FR\,I jets, but
that it is qualitatively inconsistent with the observations of 0206+35 and
0755+37 (particularly in linear polarization). 3C\,296 (modelled as
a pure outflow by \citealt{LCBH06}) may show this effect at low brightness
levels. 

$I_{\rm j}/I_{\rm cj}$ can become $< 1$ because of the effect of
aberration on anisotropic rest-frame emission.  If this acts in such a
way that the magnetic field is nearly parallel to the line of sight in
the rest frame, then the synchrotron emissivity can become very
low. If this happens in the approaching jet but not the receding one,
then the effect may be larger than that of Doppler boosting. A
symmetrical pair of jets with purely toroidal fields can show this
effect for some ranges of velocity. If the condition $\beta =
\cos\theta$ is satisfied at the edge of the approaching jet, then the
field will be exactly parallel to the line of sight in the rest frame,
so the synchrotron emissivity will be exactly zero. The condition can
never be satisfied in the receding jet (except in the trivial case of
a side-on source with zero velocity), so the sidedness ratio is also
zero. Close to the jet edge or if the velocity condition is
approximately satisfied, the sideness ratio can still be significantly
less than unity.

\begin{figure}
\begin{center}
\epsfxsize=8.5cm
\epsffile{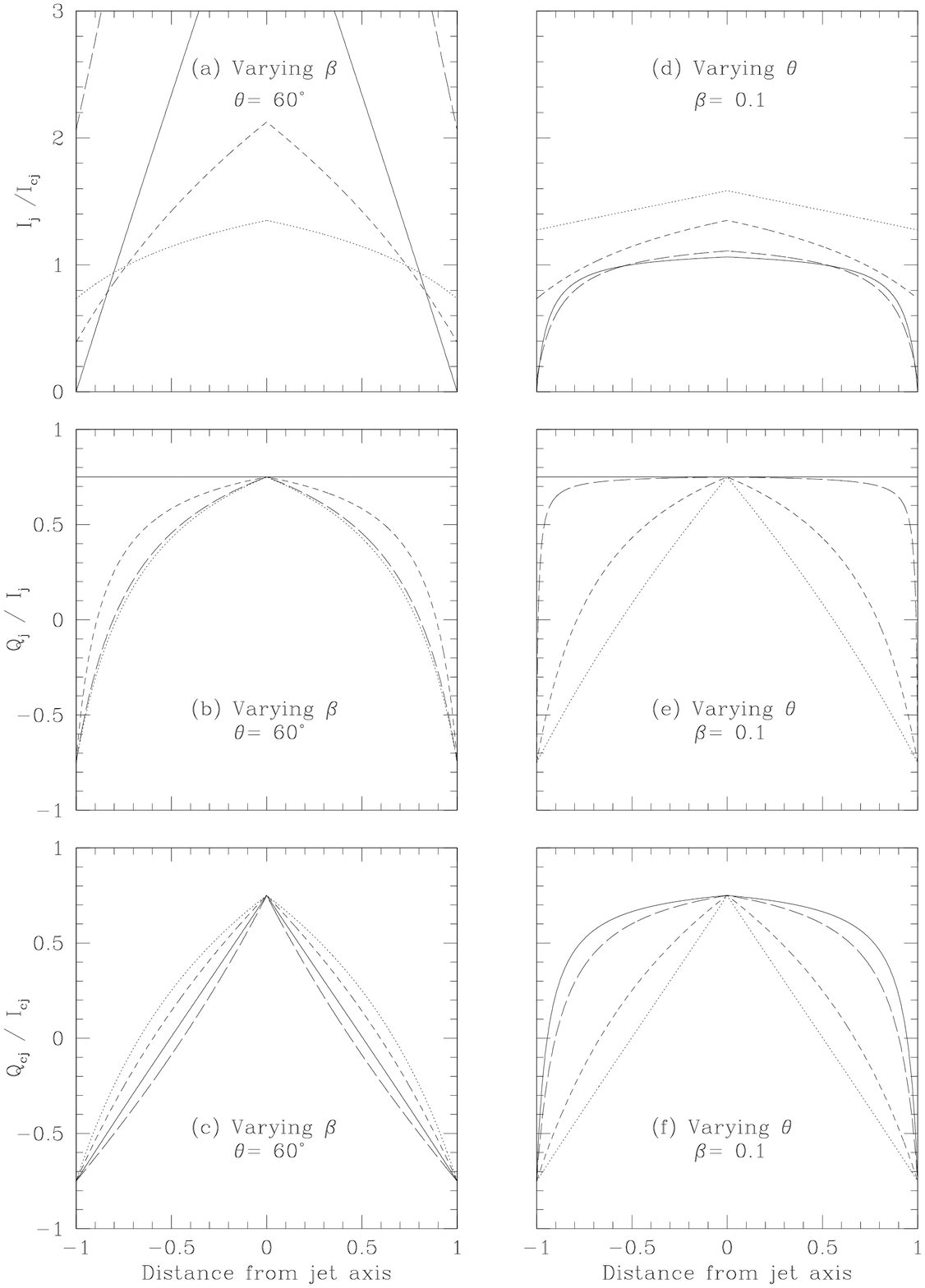}
\caption{Transverse profiles of sidedness ratio and $Q/I$ for cylindrical model
  jets containing purely toroidal fields (equations~\ref{eq:side} -- \ref{eq:dcj}). (a) and (d): sidedness ratio $I_{\rm
    j}/I_{\rm cj}$. (b) and (e): $Q_{\rm j}/I_{\rm j}$ (main jet). (c) and (f):
  $Q_{\rm cj}/I_{\rm cj}$ (counter-jet).
  (a) -- (c):  Fixed angle to the line of sight, $\theta =
  60^\circ$. The velocities are $\beta = 0.1$ (dots), 0.25 (short dash), 0.5
  (full) and 0.75 (long dash). (d) -- (f): Fixed velocity $\beta = 0.1$ The angles to
  the line of sight are $\theta = 40^\circ$ (dots), $60^\circ$ (short dash),
  $80^\circ$ (short dash) and $84.23^\circ$ (full). In all of the panels, the full
  lines represent the case $\beta = \cos\theta$, for which toroidal field loops
  are seen edge-on in the rest frame in the main jet, so $Q_{\rm j}/I_{\rm j} =
  +p_0 = +3/4$. 
\label{fig:torside}
}
\end{center}
\end{figure}

In order to demonstrate the effect, we consider a simple model with
symmetrical, cylindrical, constant-velocity jets containing purely toroidal
fields. We also take the magnetic field and radiating particle density to be
constant and assume $\alpha = 1$ so that the calculated emission profiles are
analytical, as given in the non-relativistic case by \citet{L81}. Suppose that
$x$ is a coordinate in the plane of the sky perpendicular to the projected jet
axis and normalized by the jet radius. Then the transverse profiles of sidedness
and $Q/I$ are given by:
\begin{equation}
\frac{I_{\rm j}(x)}{I_{\rm cj}(x)} = \left (\frac{D_{\rm j}}{D_{\rm
    cj}}\right )^3 \frac{(1-x^2)^{1/2} - D_{\rm j}^2\sin^2\theta
    |x|\arccos|x|}{(1-x^2)^{1/2} - D_{\rm cj}^2\sin^2\theta |x|\arccos|x|}\label{eq:side}
\end{equation}
\begin{equation}
\frac{Q_{\rm j}(x)}{I_{\rm j}(x)} = \frac{3}{4} \frac{(1-x^2)^{1/2} -
  (2-D_{\rm j}^2\sin^2\theta)|x|\arccos|x|}{(1-x^2)^{1/2} - D_{\rm j}^2\sin^2\theta
    |x|\arccos|x|}\label{eq:qij}
\end{equation}
\begin{equation}
\frac{Q_{\rm cj}(x)}{I_{\rm cj}(x)} = \frac{3}{4} \frac{(1-x^2)^{1/2} -
  (2-D_{\rm cj}^2\sin^2\theta)|x|\arccos|x|}{(1-x^2)^{1/2} - D_{\rm cj}^2\sin^2\theta
    |x|\arccos|x|}\label{eq:qicj}
\end{equation}
where the Doppler factors for the approaching and receding jets are
\begin{eqnarray}
D_{\rm j} & = & [\Gamma(1-\beta\cos\theta)]^{-1} \label{eq:dj}\\
D_{\rm cj} & = & [\Gamma(1+\beta\cos\theta)]^{-1} \label{eq:dcj}\\ \nonumber
\end{eqnarray}
We show some example profiles in Fig.~\ref{fig:torside}. We have established
that the magnetic-field structures of FR\,I jets tend to be toroidally dominated
at large distances from the nucleus and Fig.~\ref{fig:torside} shows that
$I_{\rm j}/I_{\rm cj} < 1$ at the edges for plausible velocities and angles to
the line of sight, so it would not be surprising to see this effect in some
sources. An inevitable corollary, however, is that the degree of polarization at
the edges of the jets must be high, with the apparent magnetic field transverse
to the jet axis. The reason is that the toroidal field loops are seen close to
edge-on in the rest frame: in particular, we should not observe the transition
from transverse apparent field on-axis to longitudinal at the edges. The main
jet transverse profiles of $Q/I$ for 0206+35 and 0755+37
(Figs~\ref{fig:0206trans} and \ref{fig:0755trans}) indicate that the apparent
field is primarily longitudinal ($Q/I < 0$) at the edges and certainly
inconsistent with the predicted $Q/I \approx +0.7$.  The sources must also have
$\theta \la 40^\circ$ in order to produce the large values of $I_{\rm
j}/I_{\rm cj}$ observed for their jet bases. This in turn requires high
edge velocities to satisfy the condition $\beta \approx \cos\theta$, giving a
very narrow edge with $I_{\rm j}/I_{\rm cj} < 1$.  Detailed modelling confirms
that the predicted brightness and polarization distributions are quite unlike
those observed in 0206+35 and 0755+37.

The sidedness and $Q/I$ profiles 
are, however, 
qualitatively similar to those observed in
3C\,296 \citep{LCBH06}, except that the observed value of $Q/I$ for the main jet
of 3C\,296
is $\approx$0.3, compared with the predicted 0.7 for a pure toroidal
field. Detailed modelling confirms that a simple field configuration of this
type cannot simultaneously fit the sidedness ratio and polarization, but the
edge emission is very faint 
so contamination by lobe emission
may be significant: deeper
observations are needed in order to separate jet and lobe emission unambiguously.

\end{document}